\documentclass[tighten,preprint2,trackchanges]{aastex61}

\usepackage{graphicx}
\usepackage{natbib}

\newcommand{\mum}{\ifmmode{\rm \mu m}\else{$\mu$m}\fi}
\newcommand{\iso}{{\em ISO}}
\newcommand{\iras}{{\em IRAS}}
\newcommand{\wise}{{\em WISE}}
\newcommand{\msx}{{\em MSX}}
\newcommand{\spi}{{\em Spitzer}}
\newcommand{\acet}{\ifmmode{\rm C_2H_2}\else{$\rm C_2H_2$}\fi}
\newcommand{\msmc}{{\em MSX} SMC}
\newcommand{\msun}{M$_\odot$}
\newcommand{\lsun}{L$_\odot$}
\newcommand{\rsun}{R$_\odot$}
\newcommand{\pyr}{yr$^{-1}$}

\hypersetup{colorlinks=false}

\begin{document}

\title{Characterizing the Population of Bright Infrared Sources in
 the Small Magellanic Cloud}

\author{K.~E.~Kraemer}
\affiliation{Institute for Scientific Research, Boston College, 
  140 Commonwealth Avenue, Chestnut Hill, MA 02467, USA; 
kathleen.kraemer@bc.edu}

\author{G.~C.~Sloan}
\affiliation{Center for Astrophysics and Planetary Science, Cornell 
  University, Ithaca, NY 14853-6801, USA; sloan@astro.cornell.edu}
\affiliation{Department of Physics and Astronomy, University of 
  North Carolina, Chapel Hill, NC 27599-3255, USA}
\affiliation{Space Telescope Science Institute, 3700 San Martin 
Drive, Baltimore, MD 21218, USA}

\author{P.~R.~Wood}
\affiliation{Research School of Astronomy and Astrophysics,
  Australian National University, Cotter Road, Weston Creek ACT 2611,
  Australia, wood@mso.anu.edu.au}

\author{O.~C.~Jones}
\affiliation{Space Telescope Science Institute, 3700 San Martin 
Drive, Baltimore, MD 21218, USA}

\author{M.~P.~Egan}
\affiliation{National Geospatial Intelligence Agency, 7500 GEOINT Drive, 
  Springfield, VA 22150, USA, michael.p.egan@nga.mil}


\begin{abstract}

We used {\em Spitzer}'s Infrared Spectrograph (IRS) to observe stars in the 
Small Magellanic Cloud (SMC) selected from the 
{{\em{Midcourse Space Experiment} }}
({\em{MSX}}) Point Source Catalog. We concentrate on the dust properties of 
oxygen-rich evolved stars, which show less alumina than Galactic stars. This 
difference may arise from the SMC's lower metallicity, but it could be a 
selection effect: the SMC sample includes more stars which are brighter and 
thus more massive. The distribution of SMC stars along the silicate sequence 
looks more like that of Galactic red supergiants than asymptotic giant branch 
stars (AGBs). While many are definitively AGBs, several SMC stars show 
evidence of hot bottom burning.  Other sources show mixed chemistry 
(oxygen-rich and carbon-rich features), including supergiants with PAH 
emission. MSX SMC 134 may be the first confirmed silicate/carbon star in the 
SMC, and MSX SMC 049 is a post-AGB candidate. MSX SMC 145, previously a 
candidate OH/IR star, is actually an AGB star with a background galaxy 
at $z$=0.16 along the same line-of-sight. We consider the overall 
characteristics of all the {\em MSX} sources, the most infrared-bright objects 
in the SMC, in light of {\em Spitzer}'s higher sensitivity and resolution, and 
compare them with the object types expected from the original selection 
criteria. This population represents what will be seen in more distant 
galaxies by the James Webb Space Telescope (JWST). Color-color diagrams using 
the IRS spectra and JWST mid-infrared filters show how one can separate 
evolved stars from young stellar objects (YSOs) and distinguish among 
different YSO classes.

\end{abstract}


\section{Introduction\label{sec.intro}} 

The proximity of the Small Magellanic Cloud 
(SMC) makes it possible to study in detail the various
components of a galactic ecosystem which fundamentally 
differs from our own Galaxy.  The SMC
has a comparatively low metallicity, [Fe/H] $\sim$ $-0.6$ to 
$-0.8$ \citep[][and references therein]{kw06} and
a mean distance modulus of 18.91$\pm$0.02 \citep{rubeleea15}
This gives us a nearby analog of the metal-poor systems that
populate the early Universe.

Naked stars or stars with optically thin circumstellar material 
dominate the samples observed in
the optical and near-infrared, but in the past $\sim$15--20 
years, mid-infrared maps have revealed the population of 
cooler and more embedded objects.  
To identify and characterize the various populations,
we embarked on a program to obtain infrared
spectra of representative sources with the {\it Spitzer Space 
Telescope} \citep{spitzer04} using the Infrared Spectograph 
\citep[IRS;][]{irs04}.  These spectra can test
 the population characteristics inferred from the
photometric data.

When the {\it Spitzer} mission began in 2003, the best available 
mid-infrared point source catalog of the SMC was from the 
{\it Midcourse Space Experiment} (\msx), launched by the U.S.\ Air 
Force in 1996.  On board, the Spatial Infrared Imaging Telescope 
(SPIRIT III) had four filters centered at 8.3, 12.1, 14.7, and 21.3 \mum\ 
(named A, C, D, and E, respectively). 
The astronomical observations during the cryogenic part of the mission,
described by \citet{msx01},
included surveys of the Galactic plane \citep{msx01}, the Large  Magellanic 
Cloud \citep[LMC;][]{msxlmc01}, and the SMC,
as well as Galactic star-forming regions \citep[e.g.,][]{kraemerea03},
other external galaxies \citep[e.g.,][]{kraemerea02}, and the gaps in
coverage of the {\em Infrared Astronomical Satellite}
  \citep[\iras;][]{iras84, msx01}.

The SMC catalog from \msx\ consists of the 243 sources in the main
\msx\ catalog \citep{eganea03} that lie within the region 
$ 7\fdg0\! <\! \alpha\! <\! 18\fdg7$ and $-74\fdg4\! <\! \delta \!
<\! -71\fdg5$. This catalog reveals the bright 
mid-infrared population of the SMC.  All 243 sources were detected in 
Band A, the most sensitive band, with a small
number of sources also detected in the other bands.
An additional 63 sources are in
the lower signal-to-noise ratio (SNR) catalog (SNR$\sim$3-5) with Band A 
detections and can readily be seen in the image, even though they did 
not meet the SNR criteria for the main catalog. 
The observed Band A magnitudes\footnote{The zero magnitude flux for 
 Band A of {\msx\ } is 58.5 Jy.} ranged from $\sim$3.0 to 8.2 ($\sim$0.03--3.75 
Jy), down to 8.8 mag (0.017 Jy) for the low SNR set.  In comparison, 
the other mid-IR catalog available when \spi\ launched, the \iras\
Point Source Catalog
  \citep{irascats88}
contains $\sim$50 point sources of $\sim$0.2--3 Jy in the SMC 
(including the Faint 
Source Catalog).

This paper focuses on the IRS observations of the SMC sources in
the \msx\ catalog,
with an emphasis on the objects with oxygen-rich
spectra in the infrared.
In \S 2, we describe how we selected the sources and produced 
the IRS spectra presented here.  \S 3 describes the oxygen-rich
evolved stars and their dust properties.  We assess the
photometric diagrams and discuss the results in \S 4. 
Individual sources with peculiar properties are
described in \S5. The Appendix 
describes how 
we have revised the \msmc\ catalog, using the recent 
{\it Spitzer} surveys of the SMC to improve the positional accuracy for the
source positions and 
expand our photometric knowledge of these sources.

\section{Observations and data reduction} 

\subsection{Creating the sample\label{sec.sample}} 

\begin{deluxetable*}{lllrrrrl} 
\tablecolumns{8}
\tablewidth{0pt}
\tablecaption{Observing log\label{tbl.obslog}}
\tablehead{
  \colhead{MSX} & \colhead{RA} & \colhead{Dec.} & 
  \colhead{} & \colhead{} & 
  \multicolumn{2}{c}{$t_{int}$ (sec)} & \colhead{Obs.\ date} \\
  \colhead{SMC} & \multicolumn{2}{c}{(J2000.0)} & \colhead{PID} & 
  \colhead{AOR} & \colhead{SL} & \colhead{LL} &
\colhead{(MJD\tablenotemark{a})}
}
\startdata
000      & 00 55 18.00 & $-$72 05 32.0 &  3277 & 10666240 & 240 & 1440 & 53303.5 \\
018      & 00 46 31.59 & $-$73 28 46.4 &  3277 & 10668800 & 112 &  168 & 53301.2 \\
024      & 00 42 52.23 & $-$73 50 51.7 &  3277 & 10663169 & 168 &  240 & 53521.8 \\
055      & 00 50 07.19 & $-$73 31 25.2 &  3277 & 10657536 & 112 &  240 & 53303.7 \\
067      & 00 47 36.89 & $-$73 04 44.2 &  3277 & 10664192 & 240 & 1440 & 53304.2 \\
096      & 00 50 06.40 & $-$73 28 11.2 &  3277 & 10667008 & 168 & 1200 & 53482.7 \\
109      & 00 51 29.68 & $-$73 10 44.2 &  3277 & 10667520 & 112 & 1440 & 53484.7 \\
134      & 00 50 44.40 & $-$72 37 39.0 &  3277 & 10665216 & 240 &  960 & 53304.1 \\
149      & 01 09 38.24 & $-$73 20 02.4 &  3277 & 10668032 & 112 &  960 & 53303.0 \\
168      & 00 55 26.76 & $-$72 35 56.1 &  3277 & 10668288 & 240 & 1440 & 53304.1 \\
181      & 01 00 48.17 & $-$72 51 02.1 &  3277 & 10665728 & 112 &  960 & 53303.0 \\
\\
049      & 00 44 52.56 & $-$73 18 25.9 & 30355 & 17408768 &  60 &  240 & 53914.6 \\
101      & 00 48 51.82 & $-$73 22 39.9 & 30355 & 17409024 &  60 &  360 & 53914.6 \\
145      & 00 52 12.93 & $-$73 08 53.0 & 30355 & 17409280 &  60 &  240 & 53914.6 \\
161      & 01 08 10.32 & $-$73 15 52.4 & 30355 & 17409536 &  60 &  240 & 53914.7\\
\enddata
\tablenotetext{a}{(JD$-$2400000.5)} 
\end{deluxetable*}

The 243 \msx\ SMC sources were matched with 
near-infrared data 
from the Two Micron All-Sky Survey \citep[2MASS;][]{2massexp, 2mass} 
assuming that the 2MASS source closest to the \msx\ 
coordinates was in fact the appropriate association. We then chose a sample 
of targets to span the range of colors in $J-K_s$ 
and $K_s-A$ that cover the expected positions of evolved stars,
based on previous efforts that used model spectra to classify 
sources in the \msx\ catalog of the LMC
\citep{msxlmc01}.  

\begin{deluxetable*}{llrclclrll} 
\tablewidth{0pt}
\tabletypesize{\footnotesize}
\tablecaption{Observed program sources\label{tbl.program}}
\tablehead{
  \colhead{MSX} & 
  & \multicolumn{2}{c}{Period} & \multicolumn{2}{c}{Optical} & \colhead{Infrared} & 
  \colhead{ }  &\colhead{R15} & \colhead{Var.}\\
  \colhead{SMC} & 
  \colhead{Source name} & \colhead{(days)} & \colhead{Ref.\tablenotemark{a}}& \colhead{Sp.\ Type} &  \colhead{Ref.\tablenotemark{b}}&
  \colhead{Sp.\ Class} & \colhead{$M_{bol}$} 
  & \colhead{Class\tablenotemark{c}}& \colhead{Type\tablenotemark{d}}
}
\startdata
000   & HV 12122             &    545  &G09 & M4-5.5 II/III &P83, G15& 1.N:          & $-$7.01  & O-EAGB & M: \\
018     & 2M J00463159      &    915  &G09 & M     &  K16   & 2.SE6xf       & $-$6.39   &O-AGB & \nodata \\
024     & HV 1375              &    400  &S11& M5 II         &W83, G15& 2.SE5xf:     & $-$6.27   &O-AGB & M \\
055     & IRAS 00483$-$7347    &   1859  &S11& M             &G98& 2.SE7f        & $-$7.54   &O-AGB\tablenotemark{c} & LC: \\
067     & HV 11262             & \nodata &    & K2.5-M3 I-Iabe&Multiple& 1.N           & $-$7.86   &RSG & LC: \\
096     & PMMR 34              & \nodata &    & K3-M1 I-Iab   &Multiple& 2.SE3u        & $-$7.70  &RSG & \nodata \\
109     & PMMR 41          & \nodata &    & K3-M1.5 I-Ib  &Multiple& 2.SE7u        & $-$8.03   &RSG & LC \\
134     & RAW 631              &    141  &S11& C &R93, R05, K16 & 2.SX6         & $-$5.04  &O/C-AGB\tablenotemark{c} & \nodata \\
149     & HV 2084              &    545  &P97& K5-M4 I-Iab   &Multiple& 2.SE7         & $-$8.44   &RSG & LC: \\
168     & HV 1652              & \nodata &    & K2-M0/1 I-Iab &Multiple& 2.SE8         & $-$7.79  &RSG & LC \\
181     & HV 11417             &   1092  &S11& M4-5 Ie     &E80, P83& 2.SE7         & $-$7.02  &O-AGB\tablenotemark{c} & SRc \\
049     & 2M J00445256      &    158  &S11& C & K16  & 3.C/SX        & $-$5.13  & pAGB:\tablenotemark{c} & \nodata \\
101     & PMMR 24              &    213  &P97& K2-M2 I-Iab & Multiple& 2.SE4u        & $-$7.61    &RSG & \nodata \\
145     & BMB-B 75             &    761  &S11& M6      &B80& 3.SE8+5.Uez & $-$6.28  &O-AGB+hi-z\tablenotemark{c} & SR \\
161     & IRAS F01066$-$7332   &    882  &S11& M       &G98& 3.SE7         & $-$6.02    &O-AGB & \nodata \\
\enddata
\tablenotetext{a}{Period references: P97, \cite{asas97, asas14}; 
G09, \cite{gss09}; S11, \cite{sus11} 
}
\tablenotetext{b}{Optical spectral type references: B80, \citep{bmb80}; 
E80, \cite{efh80}; P83, \cite{pmmr83}; \\
W83, \cite{wbf83}, R93,  \cite{raw93}; R05, \cite{raimondoea05};
G98, \cite{groen98}; \\
G15, \cite{gf15}; K16, this work; 
Multiple=in five or more of the references. Additional, multiply used \\
references: \cite{sk89, efh85, mo03,lev06}. 
}
\tablenotetext{c}{R15 Class: class from \cite{ruffleea15}, with the marked 
sources changed as discussed in the text.}
\tablenotetext{d}{Variability class from the GCVS
\citep{gcvs09}}
\end{deluxetable*}

Our original set of infrared spectra of \msmc\ sources was
obtained in {\it Spitzer} Cycle 1 (Program ID 3277, P.I.\ M.\ 
Egan).  This program included 35 targets from the \msmc\ 
catalog and one additional target
known to be oxygen-rich from its
optical spectrum\footnote{HV 12122 can be seen on 
the \msx\ Band A image but was below the SNR threshold for inclusion
in the catalog. This source was observed as MSX SMC 000.} \citep{smithea95}.
These targets include 19 carbon-rich stars on the 
asymptotic giant branch \citep[AGB;][]{smcc06, mcc16}, 
two R CrB candidates \citep{kra05}, one carbon-rich post-AGB object 
\citep{kra06}, and two young stellar objects \citep[YSOs;][]{jo13}.
This paper examines the remaining 11 sources in 
the sample\footnote{No data were obtained for one target due
to a typo in the input coordinates.}. Nine of these are evolved stars 
with oxygen-rich 
properties in the mid-infrared \citep[see also ][]{jonesea12,jonesea14},  
and the last two are largely featureless.  Additionally, two of the 
spectra from PID 3277 showed 
particularly strong crystalline silicate features.  
We subsequently selected four objects in the \msmc\ catalog 
with similar photometric characteristics in an effort to 
uncover additional sources with crystalline dust.
We observed these targets in {\it Spitzer} Cycle 3
(Program ID 30355, P.I.\ J.\ Houck), and three of the 
spectra  show unusual spectral features, including
 two with both oxygen- and carbon-rich features.

 Table \ref{tbl.obslog} provides particulars 
about the observations, including the coordinates from the 
photometric matching described in the appendix.
Table \ref{tbl.program}  presents the basic properties of 
the 11 oxygen-rich objects from Cycle 1 and the 4 from Cycle 3, including 
bolometric magnitudes, infrared 
spectral classifications, and object classes, which are 
explained below.

Table \ref{tbl.extra} lists 20 additional \msmc\ sources 
that were observed by other IRS programs. Overall, 59 \msmc\ sources were
observed with the IRS.
 Although \S 2 and 3 concentrate 
on the sources in Table 1, all 59 are considered in the photometric
analysis in \S 4. 
We refer the reader to the 
 papers in the table references for further details on the additional objects.

\begin{deluxetable*}{rlrrllll} 
\tablewidth{0pt}
\tablecaption{\msmc\ Targets observed in other programs\label{tbl.extra}}
\tablehead{
  \colhead{MSX} & \colhead{} & \colhead{Prog.} & \colhead{AOR} & 
  \colhead{Infrared}&  \colhead{R15}\\
  \colhead{SMC} & \colhead{Source name} & \colhead{ID}   & \colhead{key} & 
  \colhead{Sp.\ Class}  &\colhead{Class} &\colhead{$M_{bol}$}& \colhead{Ref.}
}
\startdata
027 & GM 780                &  3505 & 12931072 & 2.CE2       & C-AGB & $-$5.86 & L07 \\
039 & PMMR 23               & 50167 & 25689856 & 1.N         & RSG   & $-$7.73 & J12 \\
041 & SMP SMC 006           &   103 &  4954112 & 4.Ue        & C-PN  & $-$4.82 & BS08, BS09, S14 \\
057 & LIN 60                & 50240 & 27524608 & 5.SAuei     & YSO-1 & $-$7.24 & O13\\
072 & PMMR 19               & 50167 & 25689600 & 2.SE6       & RSG   & $-$8.41 & J12 \\
099 & S3MC 170098           & 50240 & 27518720 & 4.SAi       & YSO-1 & $-$5.71 & O13 \\
100 & SMP SMC 011           &   103 & 15902976 & 5.Ue        & C-PN  & $-$5.80 & S14 \\
102 & IRAS 00554$-$7351     &  3505 & 12936192 & 3.CE3       & C-AGB & $-$6.00 & L07 \\
104 & S3MC 170445           & 50240 & 27518208 & 4.SAi       & YSO-1 & $-$6.06 & O13 \\
107 & [MA 93] 226           & 50240 & 27524864 & 5.Eui       & YSO-1 & $-$5.72 & O13 \\
128 & ISO 00549$-$7303      &  3505 & 12932608 & 3.CE3       & C-AGB & $-$5.20 & L07 \\
133 & LHA 115-N 32 & 50240 & 27534080 & 5.Ue        & YSO-3 & $-$5.57 & LK16 \\
138 & PMMR 52               &  3505 & 12931584 & 2.SE5       & RSG   & $-$8.05 & L07 \\
153 & LIN 250               & 50240 & 27523584 & 3.SE4x       & B[e] star& $-$6.20 & R15 \\
185 & LIN 254               & 50240 & 27523072 & 4.SE3       & symbiotic& $-$5.31 & O13, R15  \\
191 & LHA 115-N 61 & 50240 & 27530240 & 5.Ue        & HII & $-$5.94 & LK16 \\
199 & IRAS 00562$-$7255     & 50240 & 27541248 & 5.Uei       & YSO-1 & $-$4.77 & O13 \\
203 & NGC 419 MIR 1         &  3505 & 12934400 & 3.CE5       & C-AGB & $-$4.92 & L07 \\
210 & HV 12149              &   200 &  6019328 & 2.SE8       & O-AGB & $-$7.03 & S08 \\
234 & LHA 115-S 38& 50240 & 27522048 & 3.SE7u:      & O-PAGB & $-$5.52 & R15 \\
\enddata
\tablerefs{
L07: \cite{lagadecea07}; J12: \cite{jonesea12};  BS08, BS09: \cite{jbs08,jbs09}; 
 S14: \cite{sloanea14};  O13: \cite{jo13};  S08: \cite{zoo08};
 LK16: Keller et al. in prep.;  R15: \cite{ruffleea15}
}
\end{deluxetable*}

\subsection{Observations} %

The spectra were observed using the low-resolution 
modules of the IRS, Short-Low (SL) and Long-Low (LL), which
provided spectra in the 5--14 and 14--37~\mum\ ranges,
respectively, at a resolution between $\sim$60 and 
120.  The spectrum in each module includes a 
second-order and first-order segment obtained in separate
apertures. The observations were made with the standard 
staring mode, which placed each target at positions one-third
and two-thirds along the slit.  Thus, a full low-resolution
spectrum combines data from eight separate pointings of
the telescope (two modules, two apertures, two nod 
positions).

For ten evolved stars with oxygen-rich dust in our Cycle 1 program,
we obtained spectra from 0.45 to 1.03~\mum\ with 
the Double-Beam Spectrograph at the 2.3 m telescope of the
Australian National University at Siding Spring Observatory.
A 0.45--0.89 \mum\ spectrum for one of the stars in program 30355
was also observed.
These spectra have a resolution of 10~\AA, and they were 
reduced with standard IRAF\footnote{Image Reduction and Analysis
Facility} procedures using the giant star 
HD~26169 to remove telluric features and HR~718 as a 
photometric standard.

\subsection{Data reduction} %

To reduce the IRS spectra, we used the Cornell spectral pipeline, which
starts with the {basic calibrated data}, 
 two-dimensional flatfielded images produced by the S18.18
reduction pipeline from the {\it Spitzer} Science Center (SSC).
The next
step differences images to remove background
emission and other additive offsets.  For the SL data,
the data with the source in the same nod position, but in 
the other aperture, served as the background image.  (This
method is known as aperture differencing.)  For LL, we used
data with the source in the opposite nod in the same aperture 
as the background image (nod differencing).  We adhered to
these rules unless background gradients or additional sources
in the slits forced us to search for a more suitable 
background image.  Differenced images were then cleaned
using the rogue pixel masks supplied by the SSC and the
{\it imclean} pixel-replacement algorithm developed at
Cornell\footnote{Available as part of the {\em irsclean}
package from the SSC.}.
  Spectra were 
extracted from the co-added images using the optimal extraction
method described by \cite{lebouteillerea10}.

We used spectral
corrections generated from observations of HR 6348 (K0 III)
for SL, and HR 6348 and HD 173511 (K5 III) for LL,
to calibrate the data photometrically, as
described by \cite{irscal15}.  We
applied this calibration separately to each of the eight
individual spectral pieces.  When combining the spectra from 
two nods in the same order, we used a sigma-clipping 
algorithm to remove spikes or divots appearing only in one 
nod.  At this point, we recalculated the spectrophotometric 
uncertainties by comparing the spectra from the two nods.

The above steps result in four calibrated spectra, from both 
spectral orders in both modules, for each star.  To combine these, 
we multiplicatively scale the segments upwards to what is 
presumably the best-centered segment and trim the overlapping
data.  These corrections are
usually less than 10\%, but they can be larger, especially if
 a source is spatially extended compared to the 3$\farcs$6 SL 
slit.   The
corrections to MSX~SMC~049 and 161 were slightly over 15\%
in one of the two SL modules, but not the other. One
of the LL nods for MSX SMC 067 was contaminated by a red source
in the slit, so only data from the other nod were used.

\subsection{Basic stellar properties} %

Table \ref{tbl.program} includes the pulsation period and optical spectral
type if known, the infrared spectral classification and object type,
the bolometric magnitude, and the variable star type from the
General Catalog of Variable Stars \citep[GCVS;][]{gcvs09}.

The pulsation periods are largely from studies using data from
the Optical Gravitational Lensing Experiment \citep[OGLE;][]{sus11}, and 
show a 
surprising amount of variation from study to study. Some of the 
disagreement can be attributed to the variability types (the last column
in Table \ref{tbl.program}), as only two sources are Mira variables. Multiple
pulsation modes are typically present in semi-regular variables, so more
than one period can often be fit to a light curve (OGLE reports up to three).
To decide among the different periods 
\citep{groen04, gss09, sus11}, we examined the light curves with 
Vizier's online Javascript tool and selected the period that gave what 
appeared to be the
best folded light curve. Typically (in six cases), this was the primary 
period given by \cite{sus11}, but in three cases we selected that of
\cite{gss09}. For two objects without OGLE data, we used the periods 
determined by the 
All Sky Automated Survey \citep[ASAS;][]{asas97, asas14}.

The optical spectral types also vary depending on the investigator.
 Some of that is probably due to different 
 spectral resolution or details in the line 
analysis. Some of the differences are real, reflecting the changes a star
undergoes during its pulsation cycle. For example, \cite{gf15} find a 
slightly different spectral type for each of up to three observational 
epochs in their sample of RSGs. Here, we try to include the range of 
spectral types that have been found, and the table references indicate 
which studies contribute to the types for each star.

The bolometric magnitudes were calculated by integrating
through the photometry in the revised \msmc\ catalog (see the Appendix)
and the IRS data.  In some cases, the photometry extends to 
wavelengths as short as the $B$ band.  A 3600 K blackbody
is used to extrapolate to the blue of the blue-most 
photometric point, and a Rayleigh-Jeans tail is used to 
extrapolate from the red end of the IRS data.  We assume an
average distance modulus of DM=18.9 \citep{rubeleea15}.

The infrared spectral classifications follow the
Hanscom method, as described by \cite{kspw02} and revised by
\citet{zoo08, mcc16} for spectra from the IRS. In this system,
the numbered group indicates the overall color of the spectral
energy distribution in the mid-infrared, from 1 for blue sources
with stellar continua to 5 to very red sources with cool dust emission.
The letters reflect the dominant spectral features, usually SE for
silicate emission in this sample. Other types used in Table
\ref{tbl.program} include:
N (naked or dust free), C (carbon-rich), SX (crystalline silicates), and U 
(unidentified infrared).
One 
source, MSX SMC 145, shows a composite spectrum, and we have 
classified the components separately (\S 
\ref{sec.145}). 

The object types are generally based on ancillary data, not just 
the IRS spectrum. They are primarily from \cite{ruffleea15}, with 
modifications for a few objects described in \S\ref{sec.hbb} and
\S5.

\section{Results}

\subsection{Dust Properties} 

\begin{figure*} 
\includegraphics[width=6.5in]{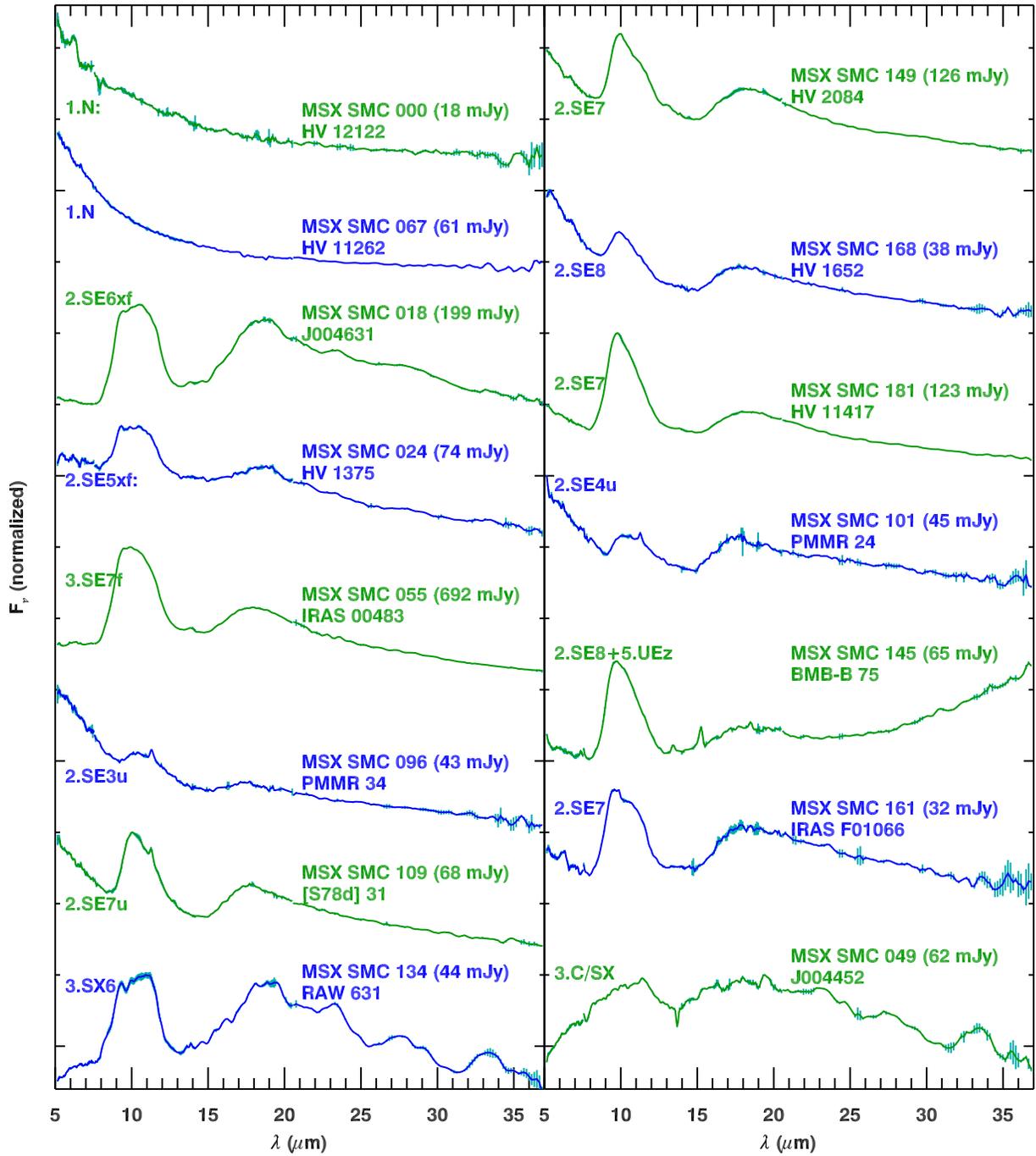} 
\caption{The IRS spectra of the 15 oxygen-rich evolved stars 
in programs 3277 and 30355. The flux in parenthesis is the equivalent 
flux for the IRAC 8 \mum\ band.} \label{fig.o.irs}
\end{figure*}

\begin{figure*} 
  \includegraphics[width=6.5in]{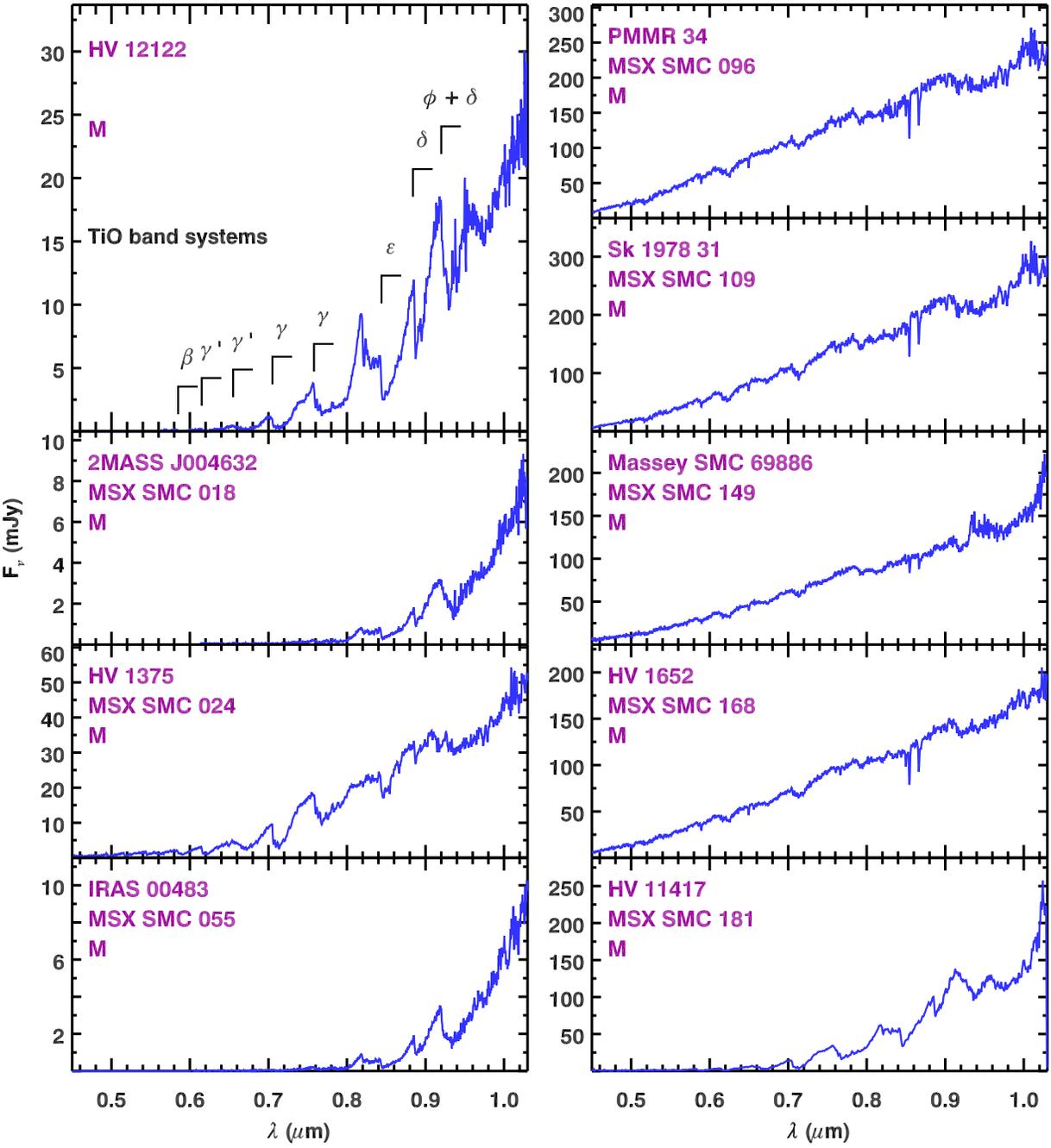}
  \caption{Optical spectra of the oxygen-rich evolved stars in
    program 3277.}
  \label{fig.o.opt}
\end{figure*}

\begin{figure} 
  \includegraphics[width=3.25in]{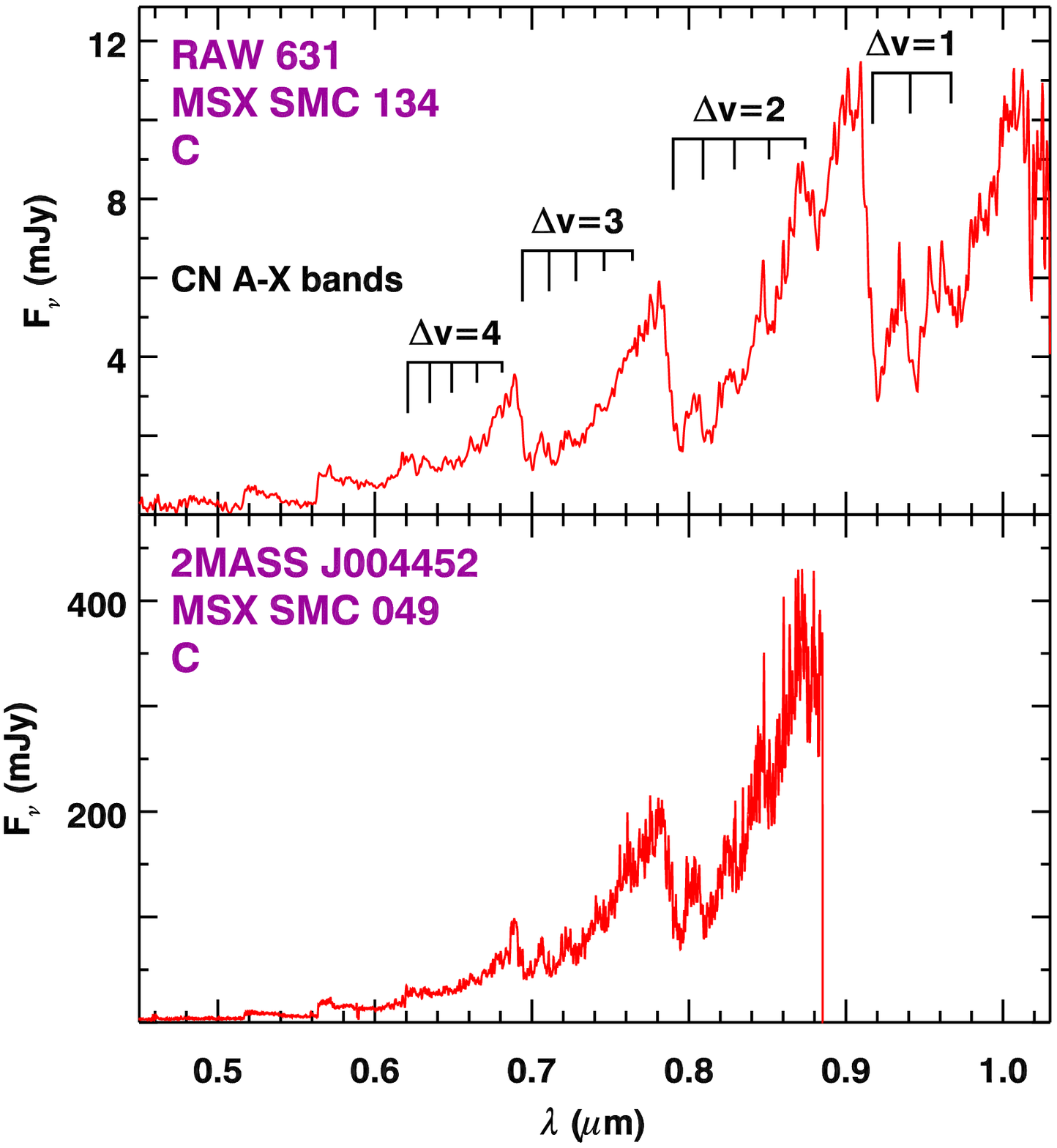}
  \caption{Optical spectra of the evolved stars in
    program 3277 and 30355 that have carbon-rich optical spectra but
    oxygen-rich features in their IRS spectra. }
  \label{fig.c.opt}
\end{figure}

Figure \ref{fig.o.irs} shows the (normalized) 
IRS spectra of the 15 sources
from programs 3277 and 30355 with infrared or
optical spectral features consistent with an oxygen-rich
chemistry, either in the dust or the photosphere. The estimated
8 \micron\ flux for Band 4 of the  Infrared Array Camera 
\citep[IRAC;][]{irac04} is given next to the name of each source.
Figures \ref{fig.o.opt} and \ref{fig.c.opt} present the optical spectra
for the sources from program 3277, as well as for MSX SMC 049. Those with
oxygen-rich photospheric features are in Figure \ref{fig.o.opt}, while
the two sources with carbon-rich features in the optical are in 
Figure \ref{fig.c.opt}.
Because MSX SMC 049 has both oxygen-rich and carbon-rich features in 
its IRS spectrum (\S \ref{sec.049}), it is excluded from the 
analysis of the oxygen-rich dust. While MSX SMC 134 has the optical
spectrum of a carbon star, its dust chemistry in the infrared is 
oxygen-rich, so it
is included here.

To characterize the properties of the oxygen-rich dust, 
we extract from the spectra several parameters that were originally 
defined by \cite{sp95, sp98} and updated by \cite{zoo08}. 
We assumed a 3600~K Planck function for the 
star, fitted to the spectrum in the 6.8--7.4~\mum\ interval.  
This continuum differs from that assumed for Galactic 
stars by \cite{sp95}, who used a 3240~K Engelke function 
\citep{eng92} with 15\% SiO absorption at 8~\mum. In the LMC and SMC,
however, \cite{zoo08} found that the SiO absorption was much weaker in 
the evolved stars.  For 
sources with strong dust features, the choice between these
two stellar continua has little impact on the dust-related 
parameters in Table \ref{tbl.o.data}.

\begin{figure} 
\includegraphics[width=3.25in]{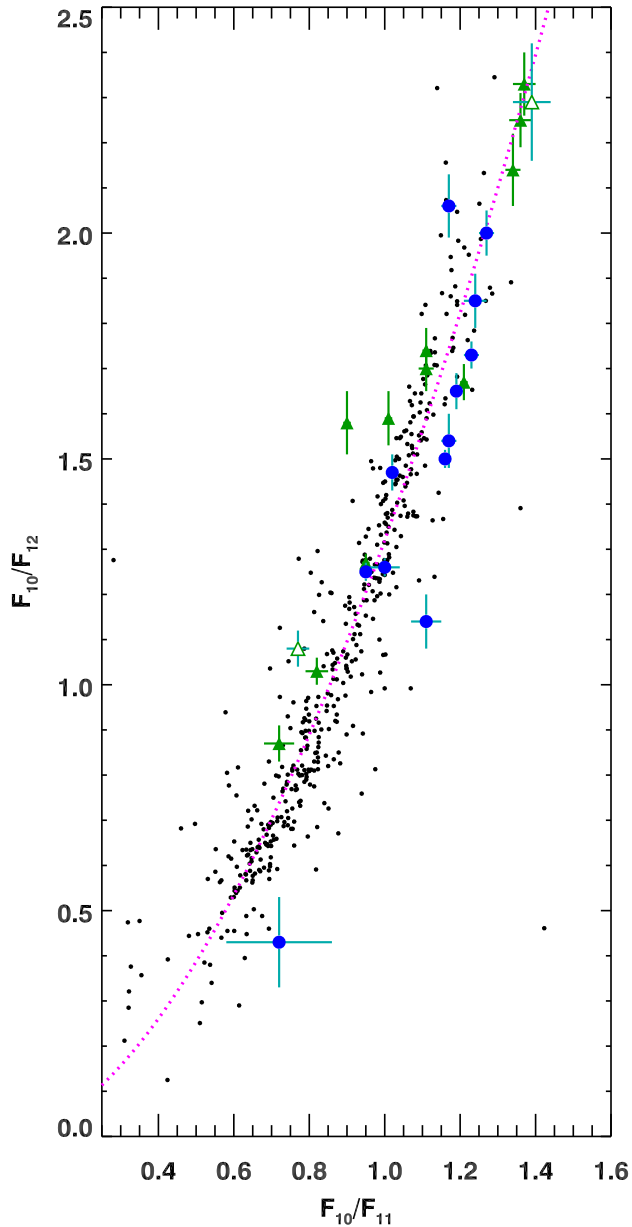}
\caption{The silicate dust sequence. The green filled triangles are the SMC 
sources 
from this work. The small black dots are
Galactic AGB sources from \cite{sp95, sp98}, the blue filled 
circles and green open triangles are the LMC  and SMC sources from 
\cite{zoo08}, 
respectively. The dotted line is the power-law fit by \cite{zoo08}.
}
\label{fig.silseq}
\end{figure}

The [7]$-$[15] color gives 
a synthetic color from two narrow bands between the dust and molecular 
features \citep{zoo08}.  Models by \cite{gss09}
have shown that the [7]$-$[15] color correlates with 
mass-loss rate, assuming similar outflow velocities and gas-to-dust 
ratios.  The remaining columns in the 
table quantify the strength and shape of the dust emission
features and  follow the definitions of 
\cite{sp95}.  We assume that the dust is
optically thin and subtract the fitted stellar continuum to
isolate the residual dust emission.  The dust emission 
contrast (DEC) is the ratio of stellar continuum to dust
excess from 7.67 to 14.03~\mum.  The strength of the dust
excess in narrow windows at 10, 11, and 12~\mum\ gives the
flux ratios $F_{10}/F_{12}$ and $F_{10}/F_{11}$.  Plotting
these ratios produces the ``silicate dust
sequence,'' which can be modeled with a simple power law (Figure 
\ref{fig.silseq}).
Shifting the flux ratios for a given spectrum to the closest
point on the silicate dust sequence gives the corrected
flux ratio $F_{11}/F_{12}$  in Table \ref{tbl.o.data}.

\begin{figure} 
\includegraphics[width=3.25in]{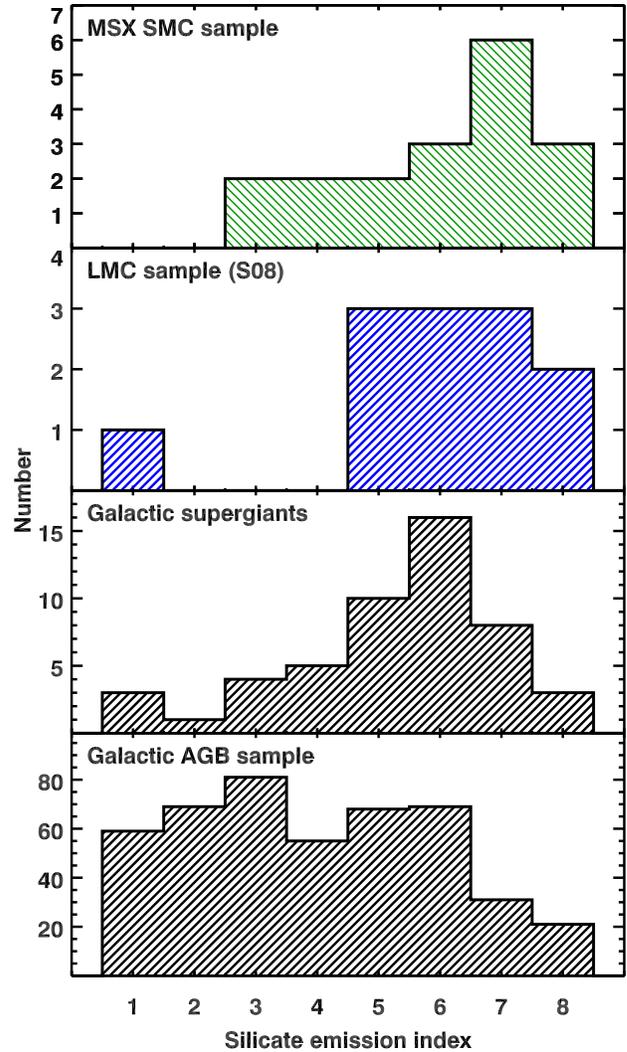}
\caption{The distribution of the silicate emission index for, from top
to bottom: the
SMC \msx\ sample, the LMC sample \citep{zoo08}, the Galactic
supergiants, and the Galactic AGB stars \citep{sp95, sp98}.
} 
\label{fig.silseqdistrib} \end{figure}

This ratio, $F_{11}/F_{12}$, in turn, leads to the silicate 
emission index\footnote{SE index $\doteq$ floor(corrected 
10*$F_{11}/F_{12}$--7.5).}
 which we use to classify the dust 
composition (last column of the table). \cite{es01} found that 
spectra dominated
by amorphous alumina-rich dust have classifications of 
SE1--3, while silicate-dominated spectra are SE6--8.  The 
SE3--6 range is more difficult to characterize; it could
arise from optically thick silicate emission starting to show
self-absorption, and/or  more
crystalline dust \citep[e.g.,][]{sdbs06, guhaniyogi11}. The SE5 and 6 sources 
in this 
sample both have 
weak crystalline features (MSX SMC 018 and 024), and one of the 
sources with strong crystallines (MSX SMC 134) is classified as
 2.SX6.

Figure \ref{fig.silseqdistrib} compares the distribution
of the SE indices for the \msx\ SMC sample (including the 6 objects in 
Table 3) with the LMC sample from 
\cite{zoo08}, and the Galactic supergiants and AGB stars from 
\cite{sp95, sp98}. These comparison samples are also shown in
the silicate dust sequence in Figure \ref{fig.silseq}.
The \msx\ sources in the SMC 
have an overabundance of SE7 and 8 sources in comparison
to both the Galactic AGB sample and the LMC sources.

It follows that spectra dominated by alumina are relatively uncommon in
the SMC sample compared to the other samples. This difference could
arise from metallicity, with the relative lack of aluminum leading to
more amorphous silicates and less alumina dust \citep[e.g.,][]{jonesea14}.
\cite{sp98} noted that the spectra of Galactic AGB stars are distributed
across the silicate sequence, whereas the RSGs are primarily SE5--7 (i.e., the
bottom two panels in Figure \ref{fig.silseqdistrib}). The SMC sample
contains both RSGs and AGB stars, but several of the AGB stars show evidence
of hot bottom burning and are likely to be massive (see \S \ref{sec.hbb}).
At a distance of 60 kpc, a bias in our sample for brighter objects would
be expected. The often uncertain distances to Galactic AGB stars make it
difficult to estimate their luminosity and hence initial mass. However,
if the trend favoring amorphous silicates in RSGs carries over to the massive
AGB stars, this could explain the lack of alumina dust in the SMC. Distances
from the {\em Gaia} mission may help address this issue in the Galaxy.

The outlier at (0.9, 1.6) in the silicate dust sequence is MSX SMC 
134, an SX6 with strong crystalline
silicate emission that enhances the $F_{12}$ measurement, and is
discussed below (\S \ref{sec.134}). One other SE6, MSX SMC 018, also has 
weak crystalline 
features. None
of the evolved stars observed here, or in any of the other programs
observing the SMC with the IRS (or with any other instrument, to the best of 
our knowledge), shows silicate features in absorption
or self-absorption. In contrast, the LMC contains several sources
with optically thick dust, both oxygen-rich \citep{zoo08} and carbon-rich
\citep{gruendlea08}. 
It is not clear if the lack of optically thick dust
in the SMC is a metallicity effect, a bias in the source selection, 
or some other effect.

With respect to the DEC,  we can point out what is a 
likely selection effect. For the present sample, the DEC ranges from 
0.19 to 3.95, with a mean and median of 1.34 and 0.96, respectively. 
As described in \S 2, our sources were selected based on their 
combined near- and mid-IR colors, with the exception of HV 12122, which 
has the smallest DEC. The four SMC spectra with silicate emission from 
\cite{zoo08} have DECs of 0.16 to 0.80,
and the 11 naked and nearly naked stars have DECs of 0.0 to 0.24. Those 
sources were chosen
based on their optical spectra. Not surprisingly, the optically selected
sample shows less dust than the infrared-selected sample.

\begin{deluxetable*}{cccccccc} 
\tablewidth{0pt}
\tablecaption{Oxygen-rich dust analysis\label{tbl.o.data}}
\tablehead{
  \colhead{MSX} & \colhead{ } & \colhead{Dust em.} & & &
  \colhead{Corrected} & \colhead{SE}\\
  \colhead{SMC} & \colhead{[7]$-$[15]} & \colhead{Contrast (DEC)} & 
  \colhead{$F_{10}/F_{11}$} & \colhead{$F_{10}/F_{12}$} &
  \colhead{$F_{11}/F_{12}$} & \colhead{Index} 
}
\startdata
HV 12122 &0.47 $\pm$0.02  & 0.19 $\pm$ 0.02 & 1.06 $\pm$ 0.03& 1.28 $\pm$ 0.04 & 1.31 $\pm$ 0.05 &\nodata\\
018      &2.03 $\pm$0.00  & 2.96 $\pm$ 0.01 & 1.01 $\pm$ 0.01& 1.59 $\pm$ 0.06 & 1.42 $\pm$ 0.05 & SE6 \\
024      &1.40 $\pm$0.01  & 0.96 $\pm$ 0.01 & 0.95 $\pm$ 0.01& 1.27 $\pm$ 0.02 & 1.29 $\pm$ 0.02 & SE5 \\
055      &1.83 $\pm$0.00  & 2.74 $\pm$ 0.01 & 1.11 $\pm$ 0.01& 1.74 $\pm$ 0.05 & 1.49 $\pm$ 0.04 & SE7 \\
067      &0.27 $\pm$0.01  & 0.02 $\pm$ 0.01 & 0.79 $\pm$ 0.17& 0.58 $\pm$ 0.13 & 0.97 $\pm$ 0.30 & \nodata\\
096      &0.62 $\pm$0.01  & 0.23 $\pm$ 0.01 & 0.72 $\pm$ 0.04& 0.87 $\pm$ 0.04 & 1.08 $\pm$ 0.07 & SE3 \\
109      &0.99 $\pm$0.01  & 0.75 $\pm$ 0.01 & 1.11 $\pm$ 0.02& 1.70 $\pm$ 0.05 & 1.47 $\pm$ 0.05 & SE7 \\
134      &1.93 $\pm$0.01  & 2.34 $\pm$ 0.03 & 0.90 $\pm$ 0.01& 1.58 $\pm$ 0.07 & 1.40 $\pm$ 0.06 & SX6 \\
149      &1.09 $\pm$0.01  & 0.90 $\pm$ 0.01 & 1.21 $\pm$ 0.01& 1.67 $\pm$ 0.04 & 1.47 $\pm$ 0.04 & SE7 \\
168      &0.65 $\pm$0.02  & 0.34 $\pm$ 0.00 & 1.34 $\pm$ 0.02& 2.14 $\pm$ 0.08 & 1.63 $\pm$ 0.06 & SE8 \\
181      &1.40 $\pm$0.01  & 1.85 $\pm$ 0.01 & 1.37 $\pm$ 0.03& 2.33 $\pm$ 0.07 & 1.69 $\pm$ 0.06 & SE8 \\
101      &0.87 $\pm$0.01  & 0.37 $\pm$ 0.01 & 0.82 $\pm$ 0.03& 1.03 $\pm$ 0.03 & 1.18 $\pm$ 0.06 & SE4 \\
145      &1.73 $\pm$0.01  & 2.09 $\pm$ 0.01 & 1.36 $\pm$ 0.03& 2.25 $\pm$ 0.06 & 1.66 $\pm$ 0.06 & SE8 \\
161      &1.60 $\pm$0.02  & 1.67 $\pm$ 0.01 & 1.10 $\pm$ 0.02& 1.83 $\pm$ 0.05 & 1.51 $\pm$ 0.05 & SE7 \\
\enddata 
\end{deluxetable*}

\subsection{AGB vs. RSG\label{sec.hbb}}

The distinction between an AGB star and an RSG often
relies on the bolometric luminosity, with $M_{bol}= -7.1$ as the 
classic limit \citep{pac71vi}. We derived the bolometric magnitudes for
the stars in Tables \ref{tbl.program} and \ref{tbl.extra} using the
IRS spectra and the photometry presented in the Appendix.
Although most, if not all, of the 
stars are variables, the luminosity derivation uses multi-epoch photometry,
so these values reflect the mean luminosity of the stars.
Three of the evolved stars
in the tables are near the classic luminosity limit, with $M_{bol} \sim -7.0$:
HV 12122, MSX SMC 181 (HV 11417), and MSX SMC 210 (HV 12149).

\cite{jonesea12} had calculated
$M_{bol}=-7.1$ for HV 12122 and called it an RSG. \cite{ruffleea15}, though,
classified it as an early-type O-rich AGB star (O-EAGB), a class of
 long-period variables (LPVs)
with weak infrared excess but no distinct dust features. As they noted,
Li absorption has been detected in the optical \citep{smithea95}, which is an
indicator of hot bottom burning (HBB). HBB occurs at the base of 
the convective envelope in the most massive AGB stars and can push them
over the classic $M_{bol}$ limit \citep[e.g.,][]{bs91,dm96, 
marigo98}.  Stars like HV 12122, with luminosities near or above the 
nominal limit and Li absorption, are mostly likely massive AGB stars 
 and HBB candidates \citep{smithea95}. In contrast, 
\cite{smithea95} did not detect Li in MSX SMC 210, so it may simply be an AGB
star relatively near the luminosity limit.

MSX SMC 181 has also been classified as both an AGB and an RSG 
\citep{ruffleea15}. \cite{efh80} first classified it as an 
M5e I based on an optical spectrum taken near maximum light, but they also 
suggest that it could be an unusual type of LPV rather 
than a real supergiant. The optical spectrum in the bottom-right panel of
Figure \ref{fig.o.opt}  shows strong TiO absorption bands, similar to the 
recognized O-AGB stars in our sample (e.g., MSX SMC 018 and 024), as opposed 
to the spectra of the RSGs (e.g., MSX SMC 096 or 168). Although listed 
as a semi-regular in the GCVS, \cite{sus11} call it a Mira due to its 
large pulsation amplitude ($\Delta I$=1.86 mag). Its luminosity has been 
estimated to be between $\sim\! -7.0$ and $-7.3$; our current estimate 
is $-7.02$. Based on its other properties, particularly the
optical spectrum and large pulsation amplitude, we call it an O-AGB, with HBB
(sometimes) pushing it above the nominal luminosity limit.

MSX SMC 055 (IRAS 00483$-$7347) has a luminosity of $-7.5$, above the
nominal limit for AGB stars. Like Hv 12122 and MSX SMC 181, though, it has 
been categorized as both an RSG and an AGB \cite[][respectively]{wfmc89, 
groen98}. Its optical spectrum \cite[Figure \ref{fig.o.opt},
see also][]{groen98} has deep TiO bands similar to the O-AGB star MSX SMC 018, 
and unlike the RSGs. Again, \cite{sus11} call it a Mira due to its large 
pulsation 
amplitude ($\Delta I$=1.73 mag) and long period (1860 d). Additionally, 
\cite{castilhoea98} detected a moderate strength Li absorption feature
at 670 nm in its optical spectrum. We conclude that this star, too,
is a high-mass AGB star where HBB allows it to exceed the classic limit.

\subsection{Hydrocarbons in RSGs\label{sec.pahs}}

\begin{figure}
\includegraphics[width=3.5in]{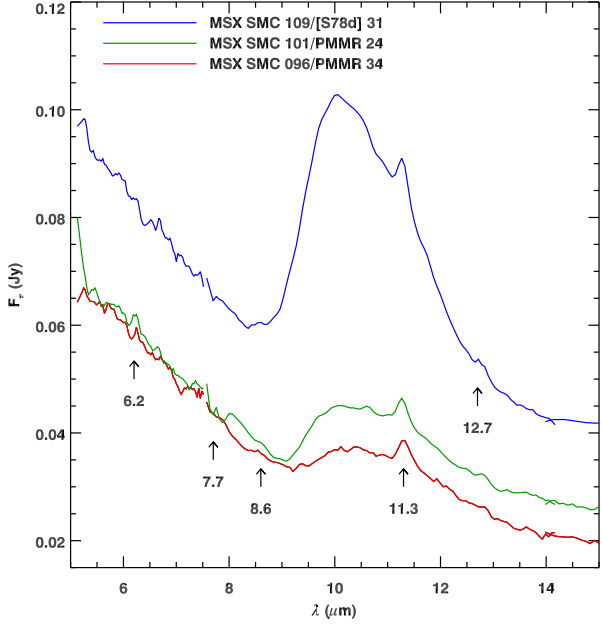}
\caption{The three RSGs with clear 11.3 \mum\ PAH emission. The 12.7 \mum\ 
feature is tentatively detected in MSX SMC 109, but none of the other 
features are convincing in it or the other RSGs.
}
\label{fig.pahs}
\end{figure}

Along with the typical silicate dust features, three of the unambiguous
RSGs in our 
sample, MSX SMC 096, 101, and 109, clearly show the 
11.3 \mum\ PAH feature (Figure \ref{fig.pahs}), as was noted by 
\cite{ruffleea15} for 096 and 109. None have strong features at the other 
expected wavelengths of PAHs, although there is a hint at
12.7 \mum\ in MSX SMC 109.
The other RSGs in Tables 2 and 3 show at most 
hints of the 11.3 \mum\ feature but are too noisy for clear detections.
This feature has previously been observed in RSGs in the Milky Way 
\citep{sylvesterea94,sylvesterea98,verhoelstea09} and the LMC 
\citep[][Jones et al., in prep.]{zoo08}. There is no extended emission 
apparent in the IRS slit images or in the IRAC images of our three RSGs 
that would indicate significant contamination in the beam. The emission 
seems to be 
localized to the RSG, just as with the other RSGs with PAH emission. 

\cite{sylvesterea98} suggested that the small fraction of RSGs in which 
they detected
PAHs implies that they arise in a short-lived phase with the needed combination
of UV flux and outflow chemistry, a combination that is not present in AGB 
stars. Another possibility is that the phenomenon is similar to the Pleiades
effect, with the RSGs embedded in diffuse material, since  \cite{lidraine02} 
found
that optical photons are sufficient to excite (pre-existing) PAHs.
This idea, however, requires that the material
containing the PAHs seen in the original RSGs in $h$ and $\chi$ Per must
be quite 
patchy within the clusters, as not all of the RSGs in those clusters show the 
feature.

\section{The Nature of the Infrared-Bright Stellar Population in the 
SMC}
\subsection{Anticipated object types\label{sec.expected}}

One of the goals of this project was to test the selection 
criteria used to build the source sample. As mentioned in \S
\ref{sec.sample}, the targets were selected based on their $J-K_s$ and $K_s-A$ 
colors, supplemented by $H-K_s$. \cite{msxlmc01} had found that that color 
space distinguished among different populations of evolved stars in 
the LMC. They used the SKY model of \cite{sky92} combined with the 
known object types for $\sim$1000 objects in the LMC. As the 
SKY model was developed from data on Galactic objects, yet seemed to 
work well in the more metal-poor environment of the LMC, it was 
expected that the SMC would show a similar agreement.

\begin{figure*}
\includegraphics[width=6.75in]{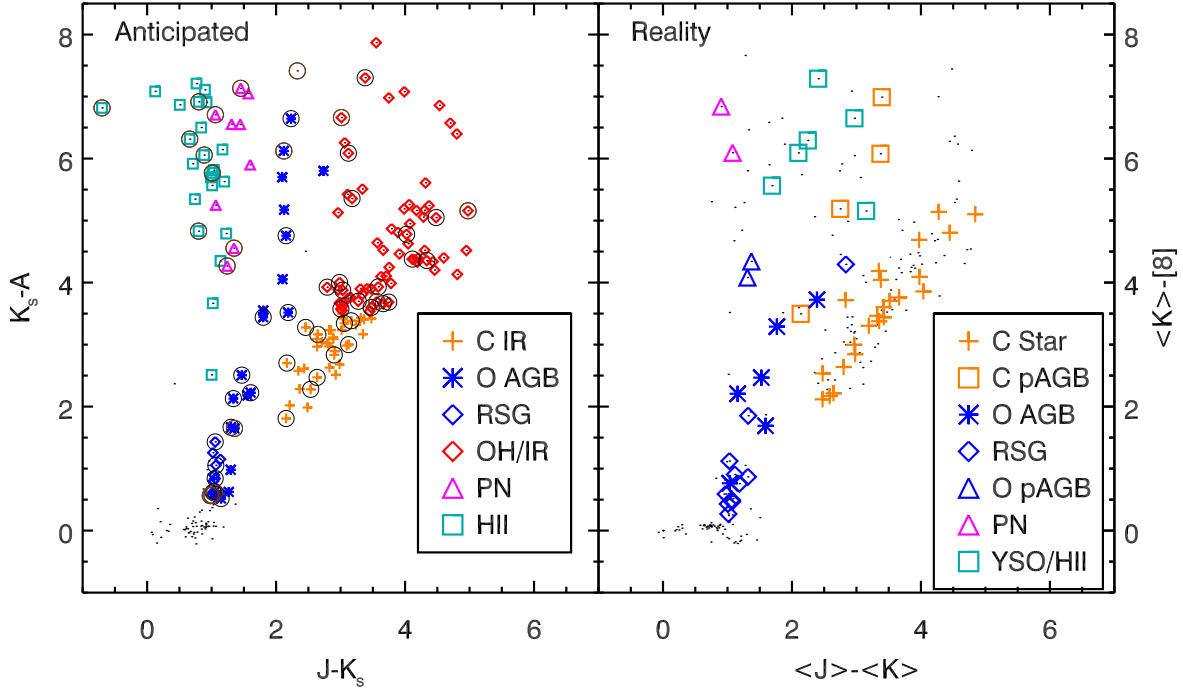}
\caption{(Left) Anticipated distribution of object types in the \msmc\ 
catalog from their \msx-2MASS colors. Circles indicate the sources that
were observed with the IRS. The clumps of dots at 
$K_s-A\sim0$ and $J-K_s\sim0-1$ are main sequence stars, and the 
small number of dots with redder colors are unclassified. (Right) The
actual object types as determined by their IRS spectra. The near-infrared
colors are from the mean photometry given in the appendix, and the 
[8] data are from IRAC Band 4. The dots here are the \msx\ SMC sources 
without IRS spectra.
}
\label{fig.befaft}
\end{figure*}

Figure \ref{fig.befaft} compares the expected types of \msmc\ objects 
based on their \msx-2MASS colors (left-hand panel) and the actual types 
based on their IRS spectra (right-hand panel). The figure includes the 
SMC sources in the \msx\ catalog that 
were observed by other programs, as well as our own. As can be seen,
more carbon-rich objects were observed than originally 
anticipated. About one-third of the proposed sample was expected to be 
carbon rich, and almost half were expected to be OH/IR stars. Instead,
two-thirds were carbon rich and no OH/IR stars were detected.
The reddest oxygen-rich objects are crystalline silicate sources, rather than 
OH/IR stars.

\cite{zijlstraea06} also found this overabundance of carbon-rich objects
in their IRS observations in the LMC. 
They suggested that the models did not well represent the objects with 
high mass-loss rates, i.e,. the redder objects. 
Furthermore,  
the oxygen-rich and carbon-rich sources overlap more in these colors 
than the SKY model 
would predict.  The carbon stars form a well-defined sequence,
but the oxygen-rich stars are more scattered. So, if one is selecting 
sources near the carbon star locus, one is likely to get a carbon-rich 
source, even when there are also oxygen-rich objects with similar colors.

\subsection{JWST color-color diagrams}

Several groups have used the modeled and observed photometry from IRAC, 
often combined with
 near-infrared and optical data, to create color-magnitude and
color-color diagrams that distinguish among types of objects such as 
YSOs, carbon stars, O-rich AGBs, etc. \citep[e.g.,][to 
name just a few]{whitneyea03, s3mc07, boyerea11}. Rapidly 
approaching, however, is the upcoming launch of the James Webb Space 
Telescope (JWST), which will host instruments with a different set of 
filters in the near- and mid-infrared. The Mid-Infrared Instrument 
\citep[MIRI;][]{miri15}, in particular, includes a set of nine filters 
that span the wavelength range from 5 to 28 \mum\ \citep{bouchetea15}, 
an excellent match to the IRS range of 5--35 \mum.

\begin{figure*}
\includegraphics[width=6.5in]{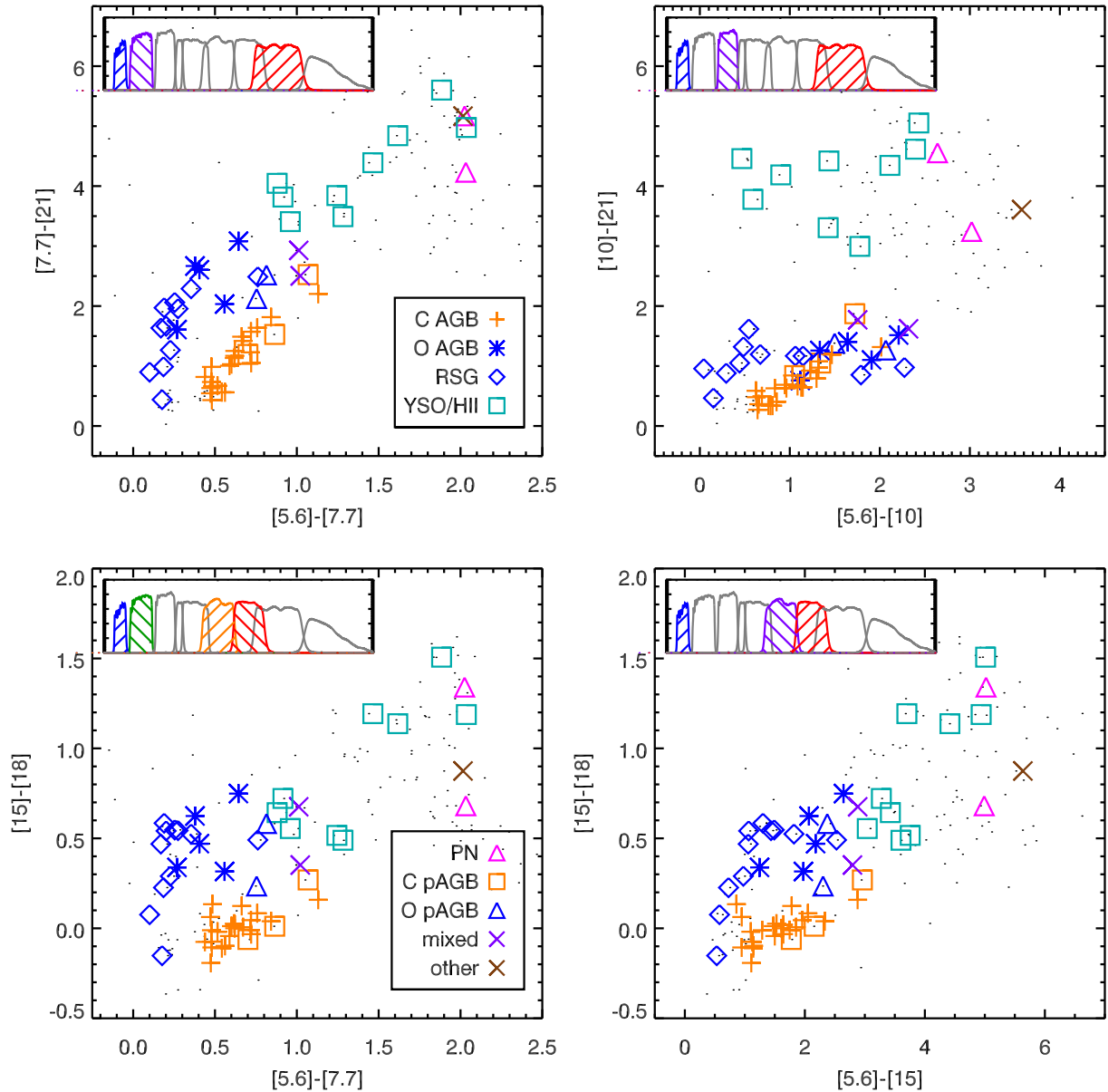}
\caption{Color-color diagrams generated from the IRS spectra using MIRI 
filter functions. The colored symbols are the \msx\ sources, with
the IRS-based object types given in the legends; the black dots are
the remaining
SMC sources with IRS spectra. (The source marked ``Other'' is MSX SMC 153, a
B[e] star.) As the MIRI filters are not yet very
familiar, we include them in the upper left of each panel, with the 
colored hatching indicating which filters are used in each plot.
\label{fig.jwstccd}
}
\end{figure*}

Although JWST will be more sensitive than \spi,  it will 
typically be observing more distant and fainter galaxies. 
The individual sources that will be detected will be the brightest, 
most luminous objects, similar to the point sources detected in the 
SMC by \msx. We can gain insight into what JWST may see by comparing
the MIRI filters to the IRS data. We have convolved the MIRI filter 
functions with the 59 IRS spectra of the \msx\ SMC sources and considered 
several combinations 
for color-color diagrams. Because the MIRI 5.6 \mum\ filter extends slightly
blueward of the IRS spectra, we extrapolate to the IRAC [4.5] photometry.
This shifts the colors slightly but makes no qualitative difference to the 
color-color diagrams here.
Figure \ref{fig.jwstccd} shows four sets with 
good separation of one or more specific object types given in Tables 
\ref{tbl.program} and \ref{tbl.extra}.

Obviously, this will not be the last word on color-color diagrams for 
JWST (see, e.g., Jones et al. submitted). A few conclusions can be drawn, 
though, from the MIRI-based diagrams. The best filters we found for 
separating the carbon stars from the O-rich AGB and RSG stars were the 
5.6, 7.7, and 21.0 \mum\ filters (upper-left panel). Most of the other 
combinations had more overlap between these types of evolved stars.

Many combinations can be used to distinguish the YSOs
from the evolved stars, as the YSOs tend to be
redder in at least one color. 
The YSOs in the bottom two panels separate into two groups. The five 
bluer sources all show silicate absorption (SA) features, and in the 
Hanscom system \citep{kspw02}, would 
be classified as 4 or 5.SAi, 
 with the ``i'' indicating ice absorption. The most prominent 
spectral feature for the redder four sources (one of which is an HII region, 
not a YSO) is PAH emission, along with fine-structure lines,
and these would be classified together as 5.UE (for UIRs + 
emission lines). 
In contrast, the SAGE-Spec classifications \citep{woodsea11, ruffleea15} 
classed two of the redder sources as YSO-1s, since an early distinguishing
feature in that decision tree is ice absorption rather than silicate 
absorption; one of the bluer sources is classed as a YSO-2, i.e., no ices.

We did not, of course, look at all 154 three- and four-filter combinations,
 let alone all those that could be generated with shorter wavelength 
filters from the Near-IR Camera \citep[NIRCam, ][]{beichmanea12}.
 Other filter combinations will undoubtedly 
highlight other categories of objects, including different SAGE-Spec YSO 
types. The [21]-[25] color, for example, could potentially be used to 
find  MgS or 
21 \mum\ features in carbon stars.

\section{Notes on Individual Sources}

\subsection{HV 12122 \& MSX SMC 067\label{sec.000}} 

HV 12122 is the source observed in program 3277 too faint
to appear in the \msmc\ catalog.  It was selected to fill out 
the samples of blue stellar sources expected to be relatively
dust free, and indeed it has one of the infrared spectra with
no obvious dust features. MSX SMC 067, aka HV 11262, also has a featureless 
IRS spectrum. However, the spectra both fall less steeply than 
Rayleigh-Jeans tails would, and they have 
measurable excesses, both in [7]$-$[15] color and in DEC.

The signal/noise ratio in the dust excess is too low to
allow a meaningful SE classification, in both
HV 12122 and MSX SMC 067.  What we have for each is a
generally featureless low-contrast excess, which could
conceivably be fitted with amorphous alumina-rich dust.
Another possible carrier is iron-rich dust, as \cite{mcd10} 
 proposed to explain low-contrast featureless
excesses in low-mass oxygen-rich AGB stars observed in
globular clusters.  

We classified HV 12122 as O-EAGB based on its optical spectrum
(Figure \ref{fig.o.opt}), which clearly shows the absorption 
band structure characteristic of TiO.  The remainder of the 
stars classified here as O-AGB were identified based on 
silicate dust emission features in their infrared spectra.
MSX SMC 067 was classified by R15 as an RSG due to its 
optical spectral type, which is supported by the bolometric magnitude that
we calculate,  $M_{bol}\sim-7.9$.

\subsection{MSX SMC 134\label{sec.134}} 

The infrared spectrum of MSX SMC 134 shows prominent emission
features from crystalline silicates at 19.5, 23.5, 
28, and 33.5~\mum, which are among the strongest that have been seen in
AGB and RSG stars \citep{jonesea12}. Even more intriguing, the
2MASS point source at this location is associated with RAW 631, a
carbon star \citep{raw93,raimondoea05}. 

Despite some extended emission in the region, the positions of the IRAC 
source, MIPS source, and 2MASS source are all within 0$\farcs$5 of each
other. Synthetic photometry from our IRS spectrum agrees with the
IRAC 8~\mum\ photometry of MSX SMC 134 within 0.26 mag 
and with the MIPS 24~\mum\ photometry within 0.05 mag. 
We have carefully examined the area around RAW 631 to ensure
that the optical and infrared spectrum are of the same 
source.  Another source lies $\sim$6\arcsec\ to the northwest
in the 2MASS images and in the Epoch 1 SAGE-SMC catalog, but  
it is 1 mag fainter at $J$ and 4 mag fainter at 
8~\mum.  No other objects appear within 10\arcsec, and
we conclude the IRS spectrum is of this source.

The optical spectrum (Figure \ref{fig.c.opt})
confirms the carbon-rich nature of the star, with strong 
absorption from the A-X CN bands throughout the optical 
region.   Van Loon et al. (2008) obtained
3--4 \mum\ spectra for a large set of IRS-targeted SMC stars, 
including MSX SMC 134. Its spectrum shows some of the weakest
acetylene (C$_2$H$_2$) and hydrogen cyanide (HCN) absorption features
in their C-rich sample (their Figures 3 and 10). The 3.1 \mum\
band is clearly detected, but the other features are not
readily apparent in the spectrum. In contrast, the mid-infrared spectrum 
is dominated by silicate dust, with no hint of carbon-rich chemistry.
Thus, the 3--4 \mum\ region examined by \cite{vanloonea08} is a transition
region between the C-rich optical emission and the silicate dust seen
in the IRS spectrum.

\cite{jonesea12}, who investigated crystalline silicates in 
the Milky Way and Magellanic Clouds,
 suggested that MSX SMC 134 is a silicate/carbon star. MSX SMC 134 would 
be the first silicate/carbon star detected
outside the Milky Way\footnote
{\cite{tramsea99} proposed that IRAS 04496$-$6958 in the LMC 
might be a silicate/carbon star, based on its spectrum from
the ISOCAM CVF on board the {\it Infrared Space Observatory}.  
It is certainly carbon rich, but its 
infrared spectrum does not show silicate emission \citep{speckea06}.  Rather,
it has some of the deepest acetylene absorption ever 
detected in the 12--16~\mum\ region, and this, along with the 
cutoff wavelength of the CVF, conspired to mislead them.}. 
The current hypothesis is that 
the oxygen-rich dust, a remnant of previous mass loss while the star was
oxygen-rich, is stored in a disk around a binary system, where it
could survive long enough for detection \citep{le90,le95},
and this remains the most likely explanation 
\citep{ydd00, le10}. 
The strong crystalline silicate features in the 
spectrum are also consistent with the notion of a disk 
\cite[e.g.,][]{molsterea99, gielenea08, gielenea11}.  
Dust grains in a disk will remain warm,
and they can be photo-processed, annealing the grains from
an amorphous to a more crystalline structure.

\begin{figure} %
\includegraphics[width=3.25in]{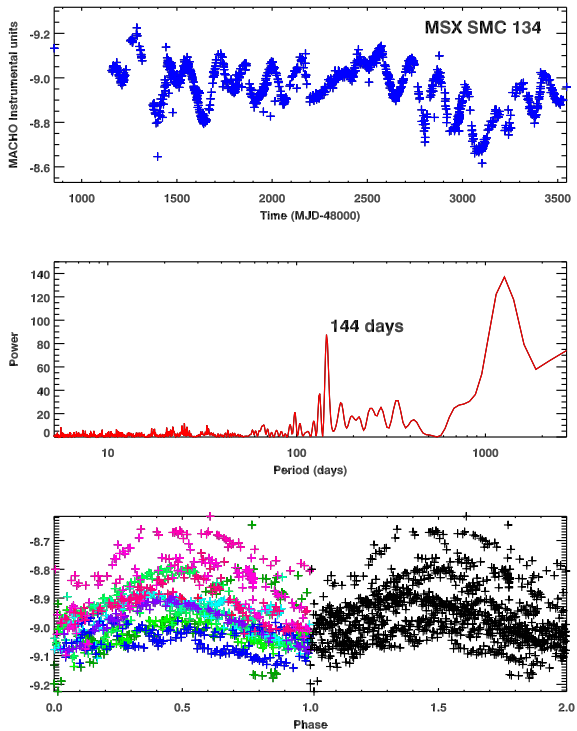}
\caption{(Top) The red-band MACHO data for MSX SMC 134. 
(Middle) The periodogram results, showing a strong peak at 144 days.
(Bottom) The data folded about the 144-day period; two phases are shown 
for clarity. The colors show sets of 144 days of data.
}
\label{fig.134} \end{figure}

MSX SMC 134 stands out among the previously known silicate/carbon stars 
 in two ways. First, 
most show little or no variability \cite[e.g.,][]{le90,kwonsuh14}
The MACHO observations of 
MSX SMC 134, though, show a clear variability (Figure
\ref{fig.134}) with two periods of
$\sim$1260 and 144 days. These are in good agreement with previous
investigations using MACHO or OGLE data by
\citet[1255 and 145 days;][]{raimondoea05} and \citet[141 days;][]{sus11},
respectively.

Second, its crystalline silicates are in
contrast with the more usual amorphous silicate features at 9.7 and 18 
\mum\ seen in most other silicate/carbon stars. There
are three others that show crystalline structure: the two post-AGB
objects HD 44179 (the Red Rectangle) and EP Lyr, as well as 
IRAS 09425$-$6040. The latter has sometimes been considered as a post-AGB 
star but is in the ``disqualified'' list of \cite[][i.e., stars that have
M spectral types or are C-AGB stars]{szczerbaea07}. MSX SMC
134 may be more like these objects than the typical 
silicate/carbon
star (although with only a few dozen known, the whole category is  
rare).

\subsection{MSX SMC 049\label{sec.049}}

\cite{ruffleea15} classified MSX SMC 049 as an O-AGB star based primarily 
on the
crystalline silicate features in the 20--35 \mum\ range of its IRS spectrum.
However, it also has a very deep acetylene absorption feature (Q branch) at 
13.7 \mum\ and  
a possible (weak) silicon carbide emission feature. In the optical,
\cite{raimondoea05} and \cite{sus11} photometrically
classified it as a carbon star. Our optical spectrum  confirms the 
carbon-rich chemistry of the source (Figure \ref{fig.c.opt}). 

\begin{figure}[h!]
\begin{center}
\includegraphics[width=3.5in]{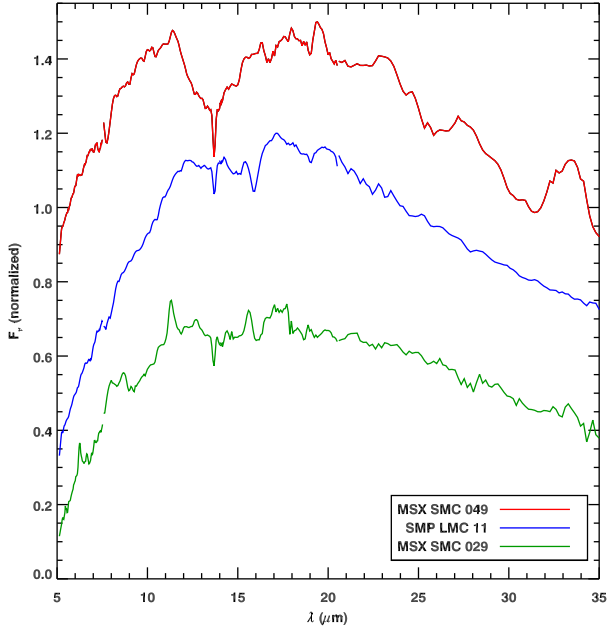}
\caption{Comparison of MSX SMC 049 (red) with SMP LMC 11 (blue) and MSX
SMC 029 (green).}
\label{fig.049}
\end{center}
\end{figure}

 The overall shape of the IRS spectrum of this mixed chemistry source
is reminiscent of the Magellanic post-AGB stars SMP LMC 
11 \citep{jbs06} and  MSX SMX 029 \citep{kra06}, shown in Figure 
\ref{fig.049}. The more complex hydrocarbons that were detected in those
objects (e.g., polyacetelynes and HC$_x$N in SMP LMC 11, PAHs and (tentatively)
 polyacetelynes in MSX SMC 029) are not as readily apparent in MSX SMC 049,
but there is a hint of absorption at 15 \mum\ that could be the $\rm 
C_6H_6+HC_3N$ feature. The shoulder around 14 \mum\ could be HCN absorption, 
and the broad absorption blueward of 13.7 might be the \acet\ R branch.

We extracted the \acet\ at 7.5
and 13.7 \mum, as well as the [6.4]--[9.3] color and silicon carbide features
using the same wavelength ranges as \cite{smcc06}. The
equivalent width of the 7.5 \mum\ feature is relatively weak, and the
13.7 \mum\ feature is not as strong as anticipated from the spectrum. The 
[6.4]--[9.3] color, though, is as red as expected, and the silicon carbide 
strength is similar to the weak SiC sources of \citet[][their Table 5]{smcc06}.
Given the apparent strength of the 13.7 \mum\ feature in the spectrum, which 
probably 
contains the R branch, a more complex treatment of these features may be 
needed to 
correctly assess the carbon-rich features.

For the crystalline features, the feature-to-continuum
ratios were fit following the methods used by \cite{jonesea12}. Except
for the 19 \mum\ feature, these are unlikely to be effected by any of the
carbon-rich features at the shorter wavelengths.  In the context of the 
large sample of crystalline silicate sources 
considered by Jones et al., the 28 and 33 \mum\ features of MSX
SMC 049 are 
among the five strongest feature-to-continuum ratios, with only their 
two SMC disk candidates (one of which is MSX SMC 134) being stronger.

As with the two carbon-rich post-AGB candidates shown in Figure \ref{fig.049}, 
the dust continuum is not well fit 
 with a single temperature, but has contributions from dust at 
$T_d\sim320-550$ K. We use the models of \cite{ssm11iv} to 
fit the infrared spectral energy distribution and
 estimate the dust mass-loss rate for MSX SMC 049
to be $\dot{M}{_d}\sim4.5\times10^{-8}$ \msun~\pyr,  although these are not
ideal for mixed chemistry sources. The models also suggest
 $L\sim6.6\times10^4$ \lsun, T$_{eff}\sim$3800 K, and R$_*\sim$370 \rsun.

\subsection{MSX SMC 145\label{sec.145}}
\begin{figure} 
\includegraphics[width=3.5in]{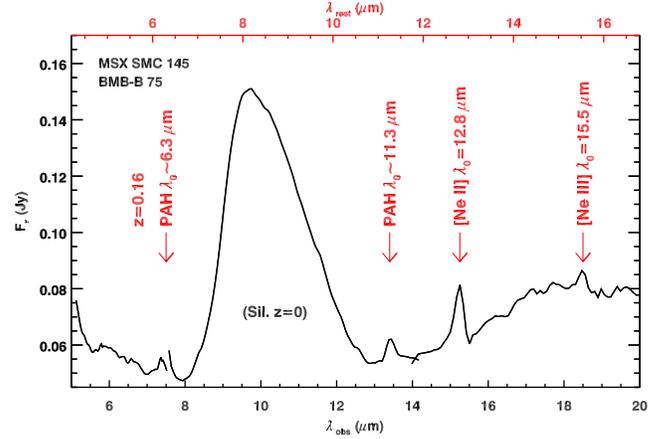}
\caption{The redshifted fine-structure lines in the spectrum of MSX SMC 145.
The 9.7 \mum\ and 18 \mum\ silicate features belong to the O-rich AGB star 
in the SMC that was the intended target.}
\label{fig.145}
\end{figure}

At first glance, the IRS spectrum of MSX SMC 145, aka BMB-B 75, 
appears to be a fairly normal silicate emission spectrum 
contaminated by a cooler dust continuum rising toward the red
and a few emission lines. Closer inspection of the location
of the narrower emission features, however, shows that they are not
at the wavelengths of any known molecular or fine-structure
lines. Instead, they appear to be redshifted emission from
a background galaxy that aligns perfectly with our intended target.
The two features at 13.4 and 15.3 \mum\ correspond to the PAH feature at
11.3 \mum\  and the [Ne II] line at 12.8 \mum\ at
a redshift of z$\sim$0.16. It then follows that the feature at 18.5 \mum\
is the 15.5 \mum\ [Ne III] line, and the 6.3 \mum\
PAH feature is tentatively detected at 7.5 \mum. Figure \ref{fig.145}
shows the 5--20 \mum\ data with the redshifted emission features labeled. 
\cite{polsdoferea15} suggested that this object might be the first OH/IR 
star found in the SMC, as it is the only star in their sample with
a broadband detection at 6 cm \citep{wongea12}, which could
potentially be from 
OH masers. \cite{jonesea15} note that MSX SMC 145 is the only
AGB star in the SMC with far-infrared emission detected by {\em Herschel}
(only one or two O-AGB stars in the LMC were detected, as well).
Based on the IRS spectrum and the redshifted fine-structure lines,
it seems more likely the radio and far-infrared emission are from the 
background galaxy, so that there 
are still no known OH/IR stars in the SMC \cite[e.g.,][]{vl12}. 

\section{Summary}

We have observed a set of infrared-bright \msx\ sources in the SMC with 
\spi's IRS and analyzed those spectra with oxygen-rich dust
features. There are relatively more sources with an index of SE7 and SE8 on 
the silicate sequence \citep{sp95,sp98} compared to LMC or Milky Way 
samples. This could be due to the number of RSGs in our sample or caused
by a metallicity-induced difference in the dust composition in the SMC.
Three of the sources with luminosities near or above the classic 
AGB/RSG boundary are 
likely still on the AGB, but with luminosities elevated due to HBB.
 
The infrared-based selection criteria
are likely responsible for the larger average DEC in our sample
compared to the SMC sources of \cite{zoo08}, who chose
their sources based on optical spectra. No deeply embedded evolved
stars of either oxygen-rich or carbon-rich chemistry have been found in 
the SMC, in contrast to the populations of OH/IR and extreme carbon
stars seen in the LMC and Milky Way. The reddest oxygen-rich sources in the
SMC are those with crystalline silicate features.
The discrepancies between the expected 
source types and those observed is likely due to how older models
treat sources with high mass-loss rates, as well as the overabundance of
carbon stars due to the lower metallicity in the SMC.

Over half of the oxygen-rich sources in our sample 
have unusual spectral features, as opposed to the LMC and Milky Way 
samples of silicate dust sources from \iras\ or \iso, which
are dominated by ordinary SE sources with no additional features. Four
sources have crystalline silicate features, including two that are among the
strongest observed by the IRS in both the LMC and SMC. 

Several sources show a mixed chemistry, with both oxygen-rich and
carbon-rich features in their optical or IRS spectra. Three of the RSGs have
the 11.3 \mum\ PAH feature but no other PAH emission in their spectra.
MSX SMC 134 is the first silicate/carbon star detected in the SMC, 
a carbon star at optical wavelengths but with strong crystalline silicate
features in the IRS range. MSX SMC 049, which is also an optical carbon star,
has both carbon-rich molecular absorption and strong crystalline silicate
features in its IRS spectrum and may be a post-AGB object. Lastly, MSX SMC 145,
which was suggested as the first OH/IR star in the SMC, is actually an 
AGB star whose IRS spectrum is contaminated by a background galaxy 
at a redshift of $z\sim0.16$.

\acknowledgments

KEK and MPE were supported in part by NASA via the Air Force 
Research Laboratory. Support for GCS was provided by NASA through Contract
Number 1257184 issued by the Jet Propulsion Laboratory (JPL),
California Institute of Technology under NASA contract 1407. OCJ acknowledges
support from NASA grant NNX14AN06G.
This research has made use of NASA's Astrophysical
Data System Bibliographic Services, the Simbad and VizieR databases
operated at the Centre de Donn\'{e}es astronomiques de
Strasbourg, and the Infrared Science Archive at the Infrared
Processing and Analysis Center, which is operated by JPL. We thank the JWST
Helpdesk and the MIRI team for making the MIRI filter functions available 
and the referee for helpful suggestions to clarify the paper.

\appendix

\section{Catalog of \msx\ Sources in the Small Magellanic Cloud}

This appendix presents a catalog based on the 243 sources 
detected in the {\it MSX} survey of the SMC, updated with positions and
photometry from more recent space-based missions and 
ground-based surveys.  {\it MSX} mapped the infrared sky from 
1996 to 1997, concentrating on the Galactic Plane 
\citep{msx01} and specific regions such as the LMC 
\citep{msxlmc01} and SMC.  While subsequent surveys of the SMC with 
{\it Spitzer} have achieved higher spatial resolution and 
sensitivity, the {\it MSX} catalog of the SMC remains a 
useful survey for multiple reasons.  First, it provides a 
flux-limited snapshot of the brightest objects in the SMC.  
Second, it makes a good comparison with the {\it MSX} catalog 
of the more metal-rich LMC \citep{msxlmc01}.  And finally, it 
has served as the basis for a number of follow-up projects, including the 
spectroscopic {\it Spitzer} survey described in this paper.

The {\it MSX} SMC Catalog is a subset of the high-latitude 
portion of the {\it MSX} Catalog, Version 2.3 
\citep{eganea03}\footnote{Available on VizieR or via Aladin 
as catalog V/114.}.  The high-latitude catalog contains 
10,168 entries with $|b| > 6\arcdeg$, and of these, 243 lie 
in the range with $ 7\fdg0 < $RA$ < 18\fdg7$ and 
$-74\fdg4 < $Dec.\ $ < -71\fdg5$.  The {\it MSX} SMC 
Catalog consists of these 243 sources, numbered sequentially 
from 1 to 243, with corresponding entry numbers in the 
high-latitude catalog of 9918 to 10160.  All 243 sources are 
detected in \msx\ Band A, but the majority have no valid 
detections in the other \msx\ bands.

\subsection{Updated Positions}
The revisions to the {\it MSX} SMC positions are based on mid-IR 
observations with {\it Spitzer} and the {\it Wide-field 
Infrared Survey Experiment} \citep[{\it WISE};][]{wise10}, 
near-IR data from 2MASS \citep{2mass,2massexp}, the Deep 
Near-IR Survey of the Southern Sky \citep[DENIS;][]{denis00},
and the IR Survey Facility \citep[IRSF;][]{irsf07}, and 
optical data from the Magellanic Clouds Photometric Survey 
\citep[MCPS;][]{mcps02} and the OGLE-III catalog of 
LPVs in the SMC 
\citep{sus11}\footnote{OGLE is the Optical Gravitional 
Lensing Experiment, and their catalogs are at 
ogledb.astrouw.edu.pl/$\sim$ogle/CVS/.}.  Most of the 
catalogs are available at the Infrared Science Archive at 
Caltech.  The exceptions are the IRSF and MCPS data, which 
we obtained via VizieR, and OGLE-III, which we obtained from their
website.

We first searched for a match to each {\it MSX} source with 
the {\it Spitzer} data.  The program Surveying the Agents
of Galactic Evolution \citep[SAGE;][]{sage06} mapped the LMC
and was expanded with the SAGE-SMC program to the SMC
\citep{sagesmc11}.  We updated the coordinates for each 
source if it was detected by the Infrared Array Camera 
\citep[IRAC;][]{irac04}, and failing that, detected at 
24~\mum\ with the Multiband Imaging Photometer for {\it
Spitzer} \citep[MIPS;][]{mips04}.  We then searched the 2MASS
catalog with the new coordinates and updated with them when we
found a match.  

In four cases (040, 083, 094, and 220), the larger {\it MSX} beam 
included two IRAC
sources with a magnitude difference of one or less at 8~\mum.
We split these four pairs into separate sources labeled 
``A'' and ``B'', giving us 247 sources in all.  Only seven of 
the \msx\ sources have no viable 2MASS counterpart.  Three have 
IRAC positions, three have MIPS positions, and one, 
MSX~SMC~032, has no point source counterpart at all.  For this one source, 
we retained its {\it MSX} position.

\begin{figure}
\plottwo{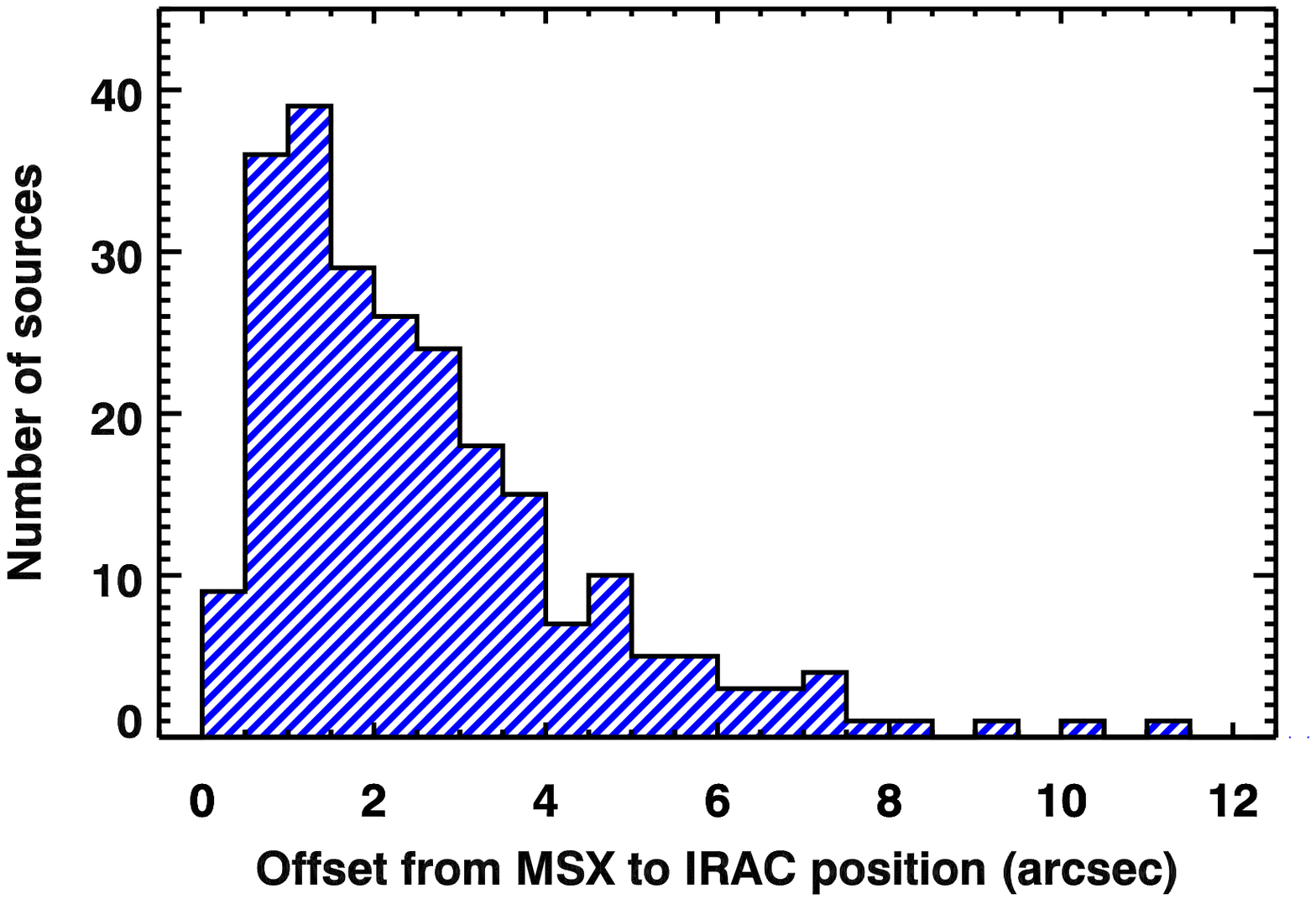}{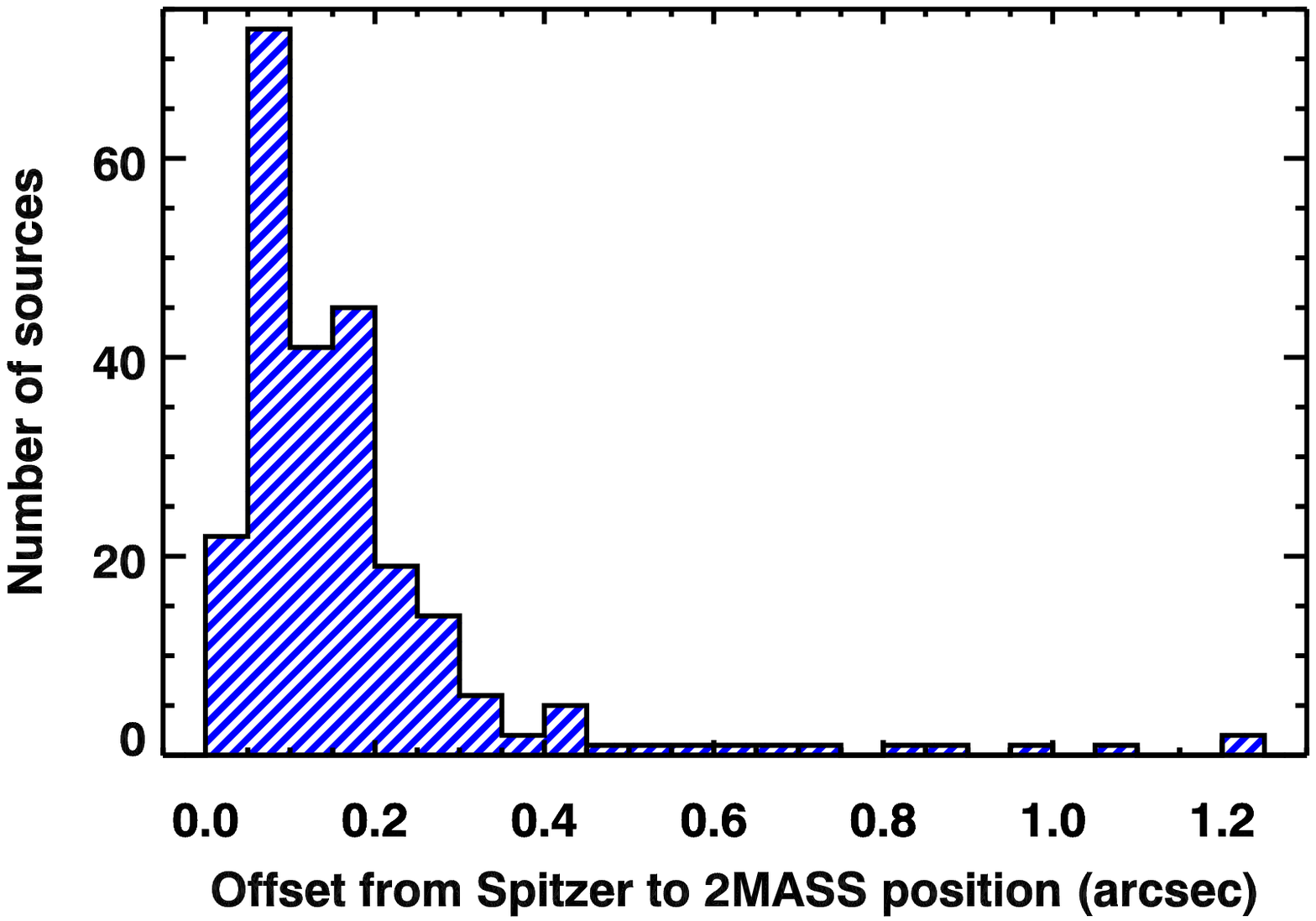}
\caption{(Left) The distributions of offsets between the 
\msx\ and \spi\ positions of the sources in the \msmc\ Catalog. 
(Right) The distributions of offsets between the new
\spi-based positions and the associated 2MASS 6x source.}
\label{fig.posns}
\end{figure}

The left panel of Figure \ref{fig.posns} compares the 
positions in the \msx\ catalog to the IRAC positions from the 
SAGE-SMC survey, and the right panel
compares the IRAC positions to 2MASS.  The median offset
between the original \msx\ positions and the corresponding IRAC
source is 2$\farcs$1, with 90\% of the sources within 
5$\farcs$0.   The median offset compares favorably to the 
quoted semi-major axis of the typical positional error 
ellipse, $\sim$2$\farcs$3 \citep{eganea03}.  The IRAC and 2MASS
positions are significantly closer, with a median separation
of 0$\farcs$12, 90\% of the sources within 0$\farcs$30, and
all of the matching sources within 1$\farcs$25.  These values
are a testament to the astrometric precision of both the
2MASS and SAGE-SMC surveys.

\begin{deluxetable*}{llrrrcrrlr}
  \tablewidth{0pt}
  \tablecaption{\msx\ SMC Source Positions\label{tab.msxpos}}
  \tablehead{
  \colhead{MSX} & \colhead{Alternate} & \colhead{RA} & \colhead{Dec.} &
  \colhead{Pos'n} & \colhead{Morph.} & \colhead{M$_{bol}$} & 
  \colhead{Obj.} & \colhead{Type} \\
  \colhead{SMC} & \colhead{Name} & \multicolumn{2}{c}{(J2000)} &
  \colhead{Ref.} & \colhead{Grade} & \colhead{(mag)} &
  \colhead{Type} & \colhead{Ref.\tablenotemark{a}}&\colhead{Notes} 
  }
  \startdata
100 &LHA 115-N 29      &12.152192  & $-$72.966919 &  2MASS&   B& $-$5.80  & PN & BS09 \\
101 &PMMR 24           &12.215924  & $-$73.377739 &  2MASS&   A& $-$7.61  & RSG & R15, K16\\
102 &IRAS 00554-7351   &14.266462  & $-$73.587410  &  2MASS&   A& $-$6.00  & C-AGB & L07 \\
103 &HD 5302           &13.276275  & $-$73.109299 &  2MASS&   A&\nodata & \nodata & \nodata & FG \ \\
104 &2MASS J00540342$-$7319384 &13.514259& $-$73.327339&  2MASS&   A&$-$6.06  & YSO & O13 \\
105 &OGLE SMC-LPV-5091 &11.258941  & $-$72.873428 &  2MASS&   A& $-$5.06  & C-AGB &S06 \\
  \enddata
\tablenotetext{a}{Object Type References: 
BS09: \cite{jbs09}; J12: \cite{jonesea12}; 
 K05: \cite{kra05}; \\
K06: \cite{kra06}; 
K16: this work; L07: \cite{lagadecea07};
O13: \cite{jo13}; R15: \cite{ruffleea15}; \\
S06: \cite{smcc06}; 
S08: \cite{zoo08}; S14: \cite{sloanea14}; 
S16: \cite{mcc16}.
}
  \tablecomments{FG: probable foreground star; CL: probable cluster member.
Table \ref{tab.msxpos} is included in its entirety
    in the electronic \\
 version of the paper. A portion is shown here for
    guidance regarding form and content. The sources shown here start with \\
MSX SMC 100 as the first several objects have nothing in the last four 
columns.}
\end{deluxetable*}
~

Table \ref{tab.msxpos} gives the positions and a reference for all 247 
sources.  It also provides a grade assessing the complexity 
of the field around each source, based on a comparison of
nearby sources in other catalogs and an examination of the 
images at 8~\mum\ (IRAC), 12~\mum\ ({\it WISE}), and 24~\mum\ 
(MIPS).  Of the 247 sources considered, 197 have a grade of
``A,'' which means that they are isolated point sources with
little or no extended emission around them.  The 36 sources 
with a ``B'' grade have an IRAC point source within 
20$\arcsec$ or extended emission which contaminates the
{\it MSX} beam.  Grades ``C'' to ``F'' describe
increasingly complex or crowded fields, with eight grade 
``C'' sources and two with grade ``D.''  Three sources 
are grade ``E'': MSX~SMC~023, 068, and 176. MSX~SMC~032, 
with no point-like
object in its vicinity at any wavelength, is the only ``F''
source.  Both MSX~SMC~023 and 032 are part of the NGC~248 LHA
115-N 13 star-forming complex.

\begin{figure*}
\includegraphics[width=6.5in]{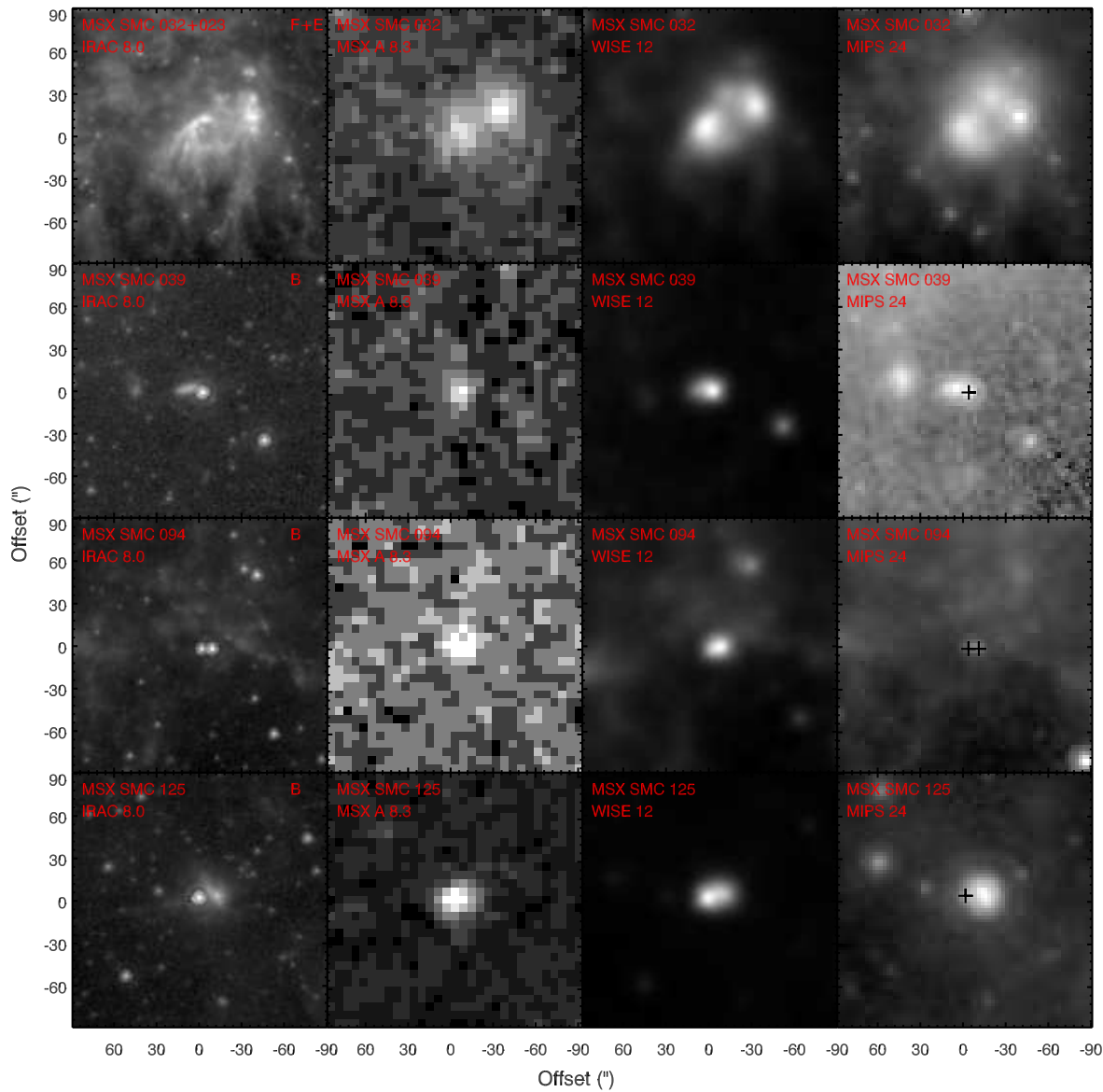}
\caption{Comparison of four \msmc\ sources with non-A grades.
The top row shows MSX SMC 032 and 023 on the left and right, grades F and E,
respectively. The lower rows show the three grade B sources described in the 
text. From left to right, the panels are IRAC 8 \mum, \msx\ 8.3 \mum,
\wise\ 12 \mum, and MIPS 24 \mum. The black plus marks in the MIPS panel
show the locations of any 2MASS point sources within 3\arcsec\ of the IRAC
position(s).}
\label{fig.mircomp}
\end{figure*}

Figure \ref{fig.mircomp} provides thumbnail sketches from {\it MSX}, IRAC,
MIPS, and {\it WISE} to illustrate some of the complex 
fields in the SMC.   In the top
row, the two ``worst'' \msx\ sources (MSX SMC 023 on the right and
032 on the left)
resolve into bright parts of NGC 248 LHA 115-N 13.
The other rows show grade B sources. For MSX SMC 039, in the second row,
an extended source is lurking to the northeast, which
contributes $\sim$25\% of the flux at 8 \mum. MSX SMC 094 has
been resolved by IRAC (and 2MASS) into two nearly equal sources. MSX SMC
125 is a carbon star, but a nearby  object ($\sim$10\arcsec), [M2002] 
SMC 21202,  contributes
about 25\% of the flux at 8 \mum\ and dominates the area at 24 \mum. The IRS
peaked up on this object rather than the intended target. Fortuitously, both
objects were in the SL slit, so those data for MSX SMC 125 could be included
in the carbon star overview study of \cite{mcc16}.

For the sources with IRS spectra, Table \ref{tab.msxpos} also includes 
the bolometric magnitude calculated by combining the spectral data
with the photometry from the next section. The type of object based on the IRS
spectrum is given, along with a reference for the classification and additional
analysis. The
last column indicates the objects that are likely foreground stars (FG), based
on having a non-zero proper motion or parallax in Simbad, or are members
of a cluster (CL), also from Simbad.

\subsection{Mean Photometry}

The mid-IR surveys provide excellent multiple-epoch coverage.  
The SAGE-SMC survey provides two epochs in all filters in 
2007--2008, and it incorporates the {\it Spitzer} Survey of 
the SMC \citep[S$^3$MC or S3MC;][]{s3mc07} which provided an 
earlier epoch for the core of the SMC in 2004.  The SAGE-VAR 
program followed up with four more epochs in the core of the 
SMC from 2010 to 2011 \citep[``VAR'' refers to the focus of 
this program on variable stars;][]{rie15}.  The {\it WISE} 
data include the reactivated mission to study near-Earth 
Objects \citep[{\em NEOWISE-R};][]{neowiser14}.  The all-sky 
scanning strategy for {\it WISE} covers the SMC every six 
months, giving two epochs in 2010 and four in the period 
2014--2015.  Thus we have up to 13 mid-IR epochs in the core
of the SMC and up to 8 in the outskirts.

Table \ref{tab.msxmir} presents the mean magnitudes in the mid-IR for each
\msx\ source.  We combined IRAC data at 3.6~\mum\ with the {\it 
WISE} data at 3.4~\mum\ (W1) without making color corrections.
We combined IRAC 4.5~\mum\ data with \wise\ 4.6~\mum\ data (W2)
similarly.  \cite{mcc16} calibrated color corrections between 
these filter pairs for carbon stars, but the present \msx\
sample contains many other source types.  The greater 
overlap between W2 and the 4.5~\mum\ IRAC filter leads to
smaller corrections compared to W1 and 3.6~\mum.

\begin{deluxetable*}{lrrrrrrr}
  \tablewidth{0pt}
  \tabletypesize{\small}
  \tablecaption{\msx\ SMC Source Photometry: Mid-IR \label{tab.msxmir}}
  \tablehead{
    \colhead{MSX} & \colhead{[3.6]\tablenotemark{a}} & \colhead{[4.5]\tablenotemark{a}} & \colhead{[5.8]}
    & \colhead{[8.0]} & \colhead{[12]} & \colhead{[22]} &
  \colhead{[24]} \\
  \colhead{SMC} & \colhead{(mag)} & \colhead{(mag)} & \colhead{(mag)} &
  \colhead{(mag)} & \colhead{(mag)} & \colhead{(mag)} & \colhead{(mag)}
  }
  \startdata
  100 & 12.469 $\pm$  0.046 & 11.922 $\pm$  0.168 & 10.033 $\pm$  0.025 &  8.188 $\pm$  0.039 &  6.419 $\pm$  0.033 &  2.527 $\pm$  0.026 &  2.353 $\pm$  0.021\\
  101 &  8.402 $\pm$  0.105 &  8.402 $\pm$  0.064 &  8.190 $\pm$  0.092 &  7.914 $\pm$  0.089 &  7.435 $\pm$  0.024 &  5.936 $\pm$  0.111 &  5.942 $\pm$  0.060\\
  102 &  9.115 $\pm$  0.441 &  8.039 $\pm$  0.398 &  7.436 $\pm$  0.024 &  6.505 $\pm$  0.028 &  5.613 $\pm$  0.021 &  4.567 $\pm$  0.028 &  4.552 $\pm$  0.404\\
  103 &  5.908 $\pm$  0.251 &  6.076 $\pm$  0.143 &  6.076 $\pm$  0.027 &  6.037 $\pm$  0.007 &  6.138 $\pm$  0.110 &  6.010 $\pm$  0.097 &  5.983 $\pm$  0.011\\
  104 & 11.195 $\pm$  0.409 &  9.149 $\pm$  0.075 &  7.867 $\pm$  0.041 &  6.820 $\pm$  0.042 &  5.642 $\pm$  0.020 &  2.481 $\pm$  0.021 &  2.290 $\pm$  0.118\\
  105 &  9.244 $\pm$  0.227 &  8.483 $\pm$  0.224 &  8.007 $\pm$  0.079 &  7.370 $\pm$  0.078 &  6.787 $\pm$  0.019 &  6.012 $\pm$  0.089 &  5.928 $\pm$  0.220\\
  \enddata
\tablenotetext{a}{No color corrections were made { when combining the IRAC and
\wise\ data  (at 3.4--3.6 \mum\ and 4.5--4.6 \mum)}.}
  \tablecomments{Table \ref{tab.msxmir} is included in its entirety
    in the electronic version of the paper. A portion is shown here for
    guidance \\
 regarding form and content. The sources start with MSX SMC 100
for consistency with Table \ref{tab.msxpos}.}
\end{deluxetable*}
~

Several epochs are also available in the near-IR.  The 2MASS
maps at $J$, $H$, and $K_s$ were obtained in 1998 August and 
followed up with a deeper survey, known as 2MASS 6X, in 2000 
December.  DENIS adds epochs at $I$, $J$, and $K_s$ from 1996 
to 1999, and the IRSF adds more at $J$, $H$, and $K_s$ from 
2002 to 2006.  Table \ref{tab.msxniropt} presents the mean magnitudes for
each filter in the near-IR.  We have not attempted to account
for subtle differences between the filter sets from the
different telescopes. The MCPS provides mean magnitudes at $U,~B,~V$, 
and $I$.  When mean
magnitudes at $V$ and $I$ are available from the LPV catalogs 
from OGLE-III, we substituted these for the MCPS values.
These mean optical and near-IR magnitudes are given in 
Table \ref{tab.msxniropt}.

\begin{deluxetable*}{lcccrrrr}
  \tablewidth{0pt}
  \tabletypesize{\small}
  \tablecaption{\msx\ SMC Source Photometry: Optical and Near-IR \label{tab.msxniropt}}
  \tablehead{
    \colhead{MSX} & \colhead{$U$\tablenotemark{a}} & \colhead{$B$} & \colhead{$V$}
    & \colhead{$I$} & \colhead{$J$} & \colhead{$H$} &
  \colhead{$K_s$} \\
  \colhead{SMC} & \colhead{(mag)} & \colhead{(mag)} & \colhead{(mag)} &
  \colhead{(mag)} & \colhead{(mag)} & \colhead{(mag)} & \colhead{(mag)}
  }
  \startdata
  100 & \nodata & \nodata & \nodata & 16.443 $\pm$  0.090 & 15.398 $\pm$  0.195 & 15.101 $\pm$  0.243 & 14.350 $\pm$    0.226\\
  101 & \nodata & \nodata & \nodata & 10.878 $\pm$  0.030 &  9.642 $\pm$  0.052 &  8.897 $\pm$  0.097 &  8.579 $\pm$    0.028\\
  102 & \nodata & \nodata & \nodata & 20.239 $\pm$  0.021 & 16.569 $\pm$  0.657 & 14.059 $\pm$  0.664 & 11.741 $\pm$    0.499\\
  103 & \nodata & \nodata & \nodata &  8.939 $\pm$  0.020 &  7.071 $\pm$  0.180 &  6.342 $\pm$  0.008 &  5.794 $\pm$    0.495\\
  104 & 18.601 $\pm$  0.074 & 18.959 $\pm$  0.048 & 18.330 $\pm$  0.111 & 17.549 $\pm$  0.132 & 16.616 $\pm$  0.283 & 15.736 $\pm$  0.193&  14.155 $\pm$  0.156\\
  105 & \nodata & \nodata & 21.099 $\pm$  9.999 & 18.370 $\pm$  0.167 & 15.151 $\pm$  0.446 & 13.082 $\pm$  0.382 &  11.245 $\pm$  0.290\\
\enddata
\tablenotetext{a}{An uncertainty of 9.999 indicates that only a single measurement was available for a given star in that band.}  
\tablecomments{Table \ref{tab.msxniropt} is included in its entirety
    in the electronic version of the paper. A portion is shown here for
    guidance \\
regarding form and content. The sources start with MSX SMC 100
for consistency with Table \ref{tab.msxpos}.}
\end{deluxetable*}
~

\bibliographystyle{aasjournal.bst}
\bibliography{orichrefs}

\end{document}